\documentclass[12pt,draft]{article}
\usepackage{psfrag,amsmath,epsfig,a4,cite,latexsym,axodraw}
\newcommand{\Eqn}[1]{Eq.~(\ref{#1})}
\newcommand{\Eqns}[2]{Eqs.~(\ref{#1}) and (\ref{#2})}

  \newcommand{\ccaption}[2]{
    \begin{center}
    \parbox{0.95\textwidth}{
      \caption[#1]{\small{#2}}
      }
    \end{center}
    }

\def\pslash#1{{\setbox0=\hbox{$#1$}
  \rlap{\ifdim\wd0>.7em\kern.22\wd0\else\kern.1\wd0\fi /}#1}}

\def\DRbar{{$\overline{DR}$}}
\def\ghat{{\hat{g}}}
\def\gtilde{{\tilde{g}}}
\def\gbar{{\bar{g}}}

\def\gth{{\breve{g}}}
\def\ath{{\breve{a}}}
\def\ttbar{t\bar{t}}

\def\ASq#1#2#3{{\cal M}_{{\rm #3}}^{#2}{(#1)}}

\def\sigmaborn{\sigma^{\rm Born}}
\def\sigmareal{\sigma^{\rm real}}
\def\sigmavirt{\sigma^{\rm virt}}
\def\sigmacoll{\sigma^{\rm coll}}

\def\MSbar{{$\overline{\mbox{MS}}$}}
\def\DRbar{{$\overline{\mbox{DR}}$}}

\def\DRED{{\scshape dred}}

\def\HV{{\scshape hv}}
\def\FDH{{\scshape fdh}}
\def\DR{{\scshape dr}}
\def\CDR{{\scshape cdr}}

\def\RS{{\scshape rs}}

\def\mDRED{{\rm\scriptscriptstyle DRED}}
\def\mCDR{{\rm\scriptscriptstyle CDR}}
\def\mHV{{\rm \scriptscriptstyle HV}}
\def\mFDH{{\rm \scriptscriptstyle FDH}}
\def\mRS{{\rm \scriptscriptstyle RS}}
\def\mFS{{\rm \scriptscriptstyle FS}}

\def\mNS{{\rm \scriptscriptstyle NS}}
\def\mMS{\overline{\rm \scriptscriptstyle MS}}

\def\eps{\epsilon}

\allowdisplaybreaks[1]
\begin{document}
\thispagestyle{empty}
\begin{flushright}
IPPP/08/54\\
DCPT/08/108\\
\end{flushright}
\vspace{3em}
\begin{center}
{\Large\bf Using Dimensional Reduction\\[1ex]
 for Hadronic Collisions}
\\
\vspace{3em}
{\sc Adrian Signer$^a$ and Dominik St\"ockinger$^b$
}\\[2em]
{\sl ${}^a$Institute for Particle Physics Phenomenology,\\ 
Durham University,
Durham DH1~3LE, UK \\
\vspace{0.3cm}
${}^b$Institut f\"ur Kern- und Teilchenphysik,\\
TU Dresden, D-01062 Dresden, Germany}
\setcounter{footnote}{0}
\end{center}
\vspace{2ex}
\begin{abstract}
We discuss how to apply regularization by dimensional reduction for
computing hadronic cross sections at next-to-leading order.  We
analyze the infrared singularity structure, demonstrate that there are
no problems with factorization, and show how to use dimensional
reduction in conjunction with standard parton distribution
functions. We clarify that different versions of dimensional reduction
with different infrared and factorization behaviour have been used in
the literature. Finally, we give transition rules for translating the
various parts of next-to-leading order cross sections from dimensional
reduction to other regularization schemes.
\end{abstract}

\vspace{0.5cm}
\centerline

\newpage
\setcounter{page}{1}

\hyphenation{
super-sym-metry-violating
coun-ter-term
coun-ter-terms
mani-fest
pho-tino
}

\section{Introduction \label{sec:introduction}}

Recently progress on the understanding of regularization by
dimensional reduction has been achieved in three directions. First, a
mathematically consistent definition avoiding the problem found in
Refs.~\cite{Siegel80,JJReview} was formulated, and a succinct method to
check the symmetry properties of dimensional reduction was developed
\cite{DS05}, leading to the verification of supersymmetry in important
cases at the two-loop level \cite{HollikDS05}. Second, explicit
calculations demonstrated how dimensional reduction can be applied to
multiloop calculations and how renormalization has to be carried out
in a non-supersymmetric context \cite{SteinhauserKantPapers}. This
provides the basis of transition rules between various definitions of
parameters such as $\alpha_s$ or $m_b$ and is useful to derive the
GUT-scale values of these parameters from the experimental values
\cite{SteinhauserTransition}. Third, an obstacle in the application of
dimensional reduction to hadronic processes was removed
\cite{Signer:2005iu} by the resolution of the factorization problem of
dimensional reduction found in Refs.~\cite{Beenakker:1988bq,
Beenakker:1996dw, Smith:2004ck}.

The purpose of the present article is to further elaborate on the
application of dimensional reduction to hadronic processes. In
Ref.~\cite{Signer:2005iu} we restricted ourselves to the case
considered in Ref.~\cite{Beenakker:1988bq}, the real corrections to
the process $gg\to \ttbar$, and showed that, despite first
appearances, in the collinear limit these real corrections factorize
into products of splitting functions and leading-order cross sections.

Here we will consider real and virtual NLO QCD corrections to
arbitrary hadronic $2\to (n-2)$ processes with massless or massive
partons. We will discuss the infrared singularity structure and the
associated regularization-scheme dependence of all these corrections,
provide transition rules between the schemes and show that all
singularities factorize. In this way we show that the framework of
dimensional reduction is completely consistent with factorization, and
we show how this scheme can be used to compute hadronic processes in
practice.

One of the main points of this article is the distinction of two
different versions of dimensional reduction that have been used in the
literature. One of the reasons why the factorization problem of
Refs.~\cite{Beenakker:1988bq,Beenakker:1996dw,Smith:2004ck} has
remained unsolved for so long is that these two versions have mainly
been applied by two different communities. The version used in
Refs.~\cite{Beenakker:1988bq,Beenakker:1996dw,Smith:2004ck} is the
same as the one defined in Refs.~\cite{Siegel79,CJN80,DS05} and is the
one mainly used in the context of supersymmetry. The version used in
Refs.~\cite{Kunszt:1993sd,Catani:1996pk,Catani:2000ef}, which was
denoted by \DR\ and is actually equivalent to the four-dimensional
helicity (\FDH) scheme \cite{FDHrefs} at one-loop, is mainly used in
the context of QCD. For the latter version, the infrared singularity
structure and transition rules have already been derived
\cite{Kunszt:1993sd,Catani:1996pk,Catani:2000ef}.
We denote these two versions by \DRED\ and \FDH. They differ in their
treatment of external particles, in a way analogous to the difference
of the ``conventional" and  ``\,'t Hooft Veltman" versions of dimensional
regularization, \CDR\ and \HV. 

In the main part of the present article we will provide results and
transition rules for all these four regularization schemes, keeping in
mind that the results for the \FDH, \CDR\ and \HV\ schemes can already
be found in Refs.~\cite{Kunszt:1993sd, Catani:1996pk, Catani:2000ef},
while the results for \DRED\ are new. The results for the infrared
singularity structure in \CDR, \HV, \FDH\ and in \DRED\ are developed
in Section~\ref{sec:RS*} and \ref{sec:DRED}, respectively. The
practical application of \DRED\ and transition rules are discussed in
Section~\ref{sec:spxs}. Appendix~\ref{app:split} provides explicit
results for all relevant splitting functions, and in
Appendix~\ref{app:examples} we provide three explicit examples of NLO
computations in \DRED.

\subsection{Elements and scheme dependences of hadronic cross sections
\label{sec:hadXS}}

We consider $n$-parton processes with up to two hadrons in the initial
state at next-to-leading (NLO) order in QCD. The partons can be either
massless quarks $q$ or gluons $g$ or massive partons such as heavy
quarks $Q$, gluinos, or squarks. In our equations we will restrict
ourselves to the most interesting case of two initial-state partons as
the simpler cases can be obtained by straightforward
modifications. The cross sections of such processes can be written as
\begin{align} 
d\sigma\big(H_1(K_1)H_2(K_2)\to a_3\ldots a_n\big) &=
\sum_{a_1,a_2} \int_0^1 dx_1f_{a_1/H_1}(x_1) \int_0^1
dx_2 f_{a_2/H_2}(x_2) \nonumber\\&\times d\hat\sigma\big(a_1(x_1K_1)
a_2(x_2K_2);a_3\ldots a_n\big),
\label{SigmaHad}
\end{align} 
where $H_{1,2}$ are the initial-state hadrons, $K_{1,2}$ their momenta
and $a_{i}$ ($i=3\ldots n$) the final-state partons. The sums run over
all possible flavours of the initial-state partons $a_{1,2}$ of the
hard partonic cross section $d\hat\sigma$, and $f_{a_i/H_i}(x)$ denote
the appropriate parton distribution functions (PDF). In the
computation of hadronic cross sections, three scheme choices have to
be made: the choice of the renormalization scheme, the factorization
scheme, and the regularization scheme.

The functional dependence of the hard cross section $d\hat\sigma$ on
input parameters like $\alpha_s$, particle masses, etc, depends on the
renormalization and the factorization schemes.  

The choice of the renormalization scheme is equivalent to a precise
definition of the input parameters entering the computation, in
particular of $\alpha_s$, particle masses and other coupling
constants. Once it has been fixed and renormalization has been carried
out accordingly, all off-shell Green functions are finite and
unambiguously defined. Changing the renormalization scheme changes
both the functional form of $d\hat\sigma$ and the numerical values of
the input parameters, such that $d\hat\sigma$ is
renormalization-scheme independent up to terms which are formally of
higher order than NLO. Common renormalization schemes are the
\MSbar-scheme for $\alpha_s$, the 
\DRbar-scheme for supersymmetric parameters, or the on-shell scheme
for masses.%
\footnote{We stress that, although the \MSbar- and
\DRbar-schemes have originally been defined with reference to specific
regularization schemes, they can be realized in the context of any
regularization scheme.}  In the following we are not concerned with
the renormalization-scheme dependence and assume that some
renormalization scheme has been fixed.

The choice of the factorization scheme is equivalent to a precise
definition of the parton distribution functions. Both the functional
form of the hard cross section and the numerical values of the parton
distribution functions depend on this choice, but this dependence
cancels in the full hadronic cross section up to terms which are
formally of
 higher order than NLO. The major part of the following
considerations is independent of the factorization scheme, and we
simply assume that some factorization scheme has been fixed. At the
end we will specialize to the important case of the
\MSbar-factorization scheme. In that case, the parton distribution
functions $f_{a_i/H_i}$ can be taken for instance from the well-known
MRST or CTEQ sets \cite{MRST,CTEQ}.

Our main focus is the influence of the regularization scheme
(\RS). After removing the regularization, all quantities appearing in
\Eqn{SigmaHad} are \RS\ independent, but the hard partonic cross
section is a sum of \RS\ dependent parts. It is commonly written as
\begin{align}
d\hat\sigma &=
d\sigmaborn_\mRS+d\sigmareal_\mRS+d\sigmavirt_\mRS+d\sigmacoll_\mRS,
\label{sigmahat}
\end{align}
where the \RS\ dependence is explicitly indicated. The lowest-order, or
Born cross section $d\sigmaborn$ is finite, and in the limit where the
regularization is removed its \RS\ dependence vanishes. The three NLO
contributions are the real and virtual corrections $d\sigmareal$,
$d\sigmavirt$ and the collinear counterterm $d\sigmacoll$, which
subtracts initial-state collinear singularities. All NLO contributions
involve collinear and/or soft singularities and depend on the \RS\ in
their finite and their divergent parts. There are no ultraviolet
singularities and associated \RS\ dependences in $d\sigmavirt$, because
these are eliminated by renormalization and by fixing the
renormalization scheme.

\subsection{Variants of dimensional regularization and dimensional
reduction \label{sec:DR}}

In all dimensional schemes space-time is continued from $4$ to $D$
dimensions, where $D=4-2\eps$ is an arbitrary complex number. In this
way momentum integrals become well-defined and ultraviolet and
infrared singularities appear as $1/\eps^k$-poles as $\eps\to0$. Gluon
fields are treated differently in dimensional regularization and
dimensional reduction. In the former, gluons are treated as
$D$-dimensional as well; in the latter, gluons are treated as
$4$-dimensional.

Both choices have certain advantages. The purely $D$-dimensional
treatment of all objects leads to simpler expressions, but it breaks
supersymmetry owing to the different number of degrees of freedom of
the gluon and the gluino. The $4$-dimensional treatment of the gluon
is better compatible with supersymmetry and it is more amenable to
helicity methods, which are commonly used to simplify QCD higher-order
computations.

In order to formulate the two schemes one needs to distinguish
three spaces:
\begin{itemize}
\item the original 4-dimensional space (4S).
\item the formally $D$-dimensional space for momenta and momentum
  integrals. This space is actually an infinite-dimensional vector
  space with certain $D$-dimensional properties \cite{Wilson,Collins},
  and is sometimes called ``quasi-$D$-dimensional space''
  (Q$D$S). The space 4S is therefore a subspace of Q$D$S.
\item the formally 4-dimensional space for e.g.\ gluons in dimensional
  reduction. This space has to be a superspace of Q$D$S in order for
  the dimensionally reduced theory to be gauge invariant. Hence it
  cannot be identified with the original 4S --- it can only be
  constructed as a ``quasi-4-dimensional space'' (Q4S) \cite{ACV,DS05}
  with certain 4-dimensional properties. In practice the distinction
  between Q4S and 4S often does not matter, but it is important in the
  definition of the different versions of dimensional reduction, see
  below, and to   avoid the inconsistency uncovered in
  Ref.~\cite{Siegel80}. 
\end{itemize}
These three spaces are characterized by their metric tensors, which we
denote by $g^{\mu\nu}$ (for Q4S), $\ghat^{\mu\nu}$ (for Q$D$S), and
$\gbar^{\mu\nu}$ (for 4S). The dimensionalities of the spaces are
expressed by the following equations:
\begin{align}
 g^{\mu\nu}g_{\mu\nu} &= 4,&
 \ghat^{\mu\nu}\ghat_{\mu\nu} &= D=4-2\eps,&
 \gbar^{\mu\nu}\gbar_{\mu\nu} &= 4.
 \label{ggDef}
\end{align} 
The following projection relations express that 4S is a subspace of
Q$D$S and Q$D$S is a subspace of Q4S:
\begin{align} 
g^{\mu\nu}\ghat_\nu{}^\rho&=\ghat^{\mu\rho},&
g^{\mu\nu}\gbar_\nu{}^\rho&=\gbar^{\mu\rho},&
\ghat^{\mu\nu}\gbar_\nu{}^\rho&=\gbar^{\mu\rho}.
\label{ggProjections}
\end{align}
It is useful to introduce the orthogonal complement to Q$D$S. This is
a $4-D=2\epsilon$-dimensional space with metric tensor
$\gtilde^{\mu\nu}$, which satisfies
\begin{align}
\label{gghatgtilde}
g^{\mu\nu}&=\ghat^{\mu\nu}+\gtilde^{\mu\nu},\\
\gtilde^{\mu\nu}\gtilde_{\mu\nu}&=4-D=2\epsilon,\\
g^{\mu\nu}\gtilde_{\nu}{}^\rho&=\gtilde^{\mu\rho},&
\ghat^{\mu\nu}\gtilde_{\nu}{}^\rho&=0,&
\gbar^{\mu\nu}\gtilde_{\nu}{}^\rho&=0.
\end{align}

Within this framework it is now possible to precisely state the
calculational rules of dimensional regularization and dimensional
reduction. Since momenta are always treated in $D$ dimensions, it only
needs to be specified how gluons (or other
vector fields) are treated. More precisely, it needs to be specified
which metric tensors are used in gluon propagator numerators and in
gluon polarization sums. Particularly important for the understanding
of factorization is the treatment of gluon polarization sums in
squared matrix elements. Without regularization, such polarization
sums can be written as
\begin{align}
\sum_{\rm pols}\epsilon^\mu\epsilon^\nu{}^* =
-\gbar^{\mu\nu}+\frac{n^\mu k^\nu+k^\mu n^\nu}{(nk)}
-\frac{n^2 k^\mu k^\nu}{(nk)^2},
\end{align}
where $k$ is the gluon momentum and $n$ is a gauge vector such that
$nk\ne0$. With regularization, the metric tensor in this polarization
is replaced by either $g^{\mu\nu}$, $\ghat^{\mu\nu}$,
$\gtilde^{\mu\nu}$, or $\gbar^{\mu\nu}$.

It is not strictly necessary to regularize all gluons. Only gluons
that appear inside a divergent loop or phase space integral
(``internal'') need to be regularized; for all other gluons
(``external'') regularization is optional. The precise definitions of
``internal/external'' in this context are as follows: ``Internal
gluons'' are defined as either virtual gluons that are part of a
one-particle irreducible loop diagram or, for real correction
diagrams, gluons in the initial or final state that are collinear or
soft. ``External gluons'' are defined as all other gluons.
 
Now, since external gluons do not have to be
treated in the same 
way as internal ones, it  is in fact possible to distinguish two
variants of each regularization. The two variants of dimensional
regularization are:
\begin{itemize}
\item \CDR\ (``conventional dimensional regularization"): Here internal
and external gluons (and other vector fields) are all treated as
$D$-dimensional.
\item \HV\ (``\,'t Hooft Veltman scheme"): Internal gluons are treated as
$D$-dimensional but external ones are treated as strictly
4-dimensional.
\end{itemize}
Note that the above definition of internal gluons in phase space
integrals is necessary for unitarity but leads 
to complications in the treatment of phase space integrals in schemes
where internal and external gluons are treated differently.
The two analogous variants of dimensional reduction are:
\begin{itemize}
\item \DRED\ (``original/old dimensional reduction"): Internal and
external gluons are all treated as quasi-4-dimensional.
\item \FDH\ (``four-dimensional helicity scheme"): Internal gluons are
treated as quasi-4-dimensional but external ones are treated as
strictly 4-dimensional.
\end{itemize}
Table \ref{tab:RSs} illustrates these four schemes.

\begin{table}
\begin{center}
\begin{tabular}{l|cccc}
&\CDR&\HV&\FDH&\DRED\\
\hline
internal gluon&$\ghat^{\mu\nu}$&$\ghat^{\mu\nu}$&
$g^{\mu\nu}$&$g^{\mu\nu}$\\
 external gluon&$\ghat^{\mu\nu}$&$\gbar^{\mu\nu}$&
$\gbar^{\mu\nu}$&$g^{\mu\nu}$
\end{tabular}
\end{center}
\ccaption{}{
Treatment of internal and external gluons in the four different \RS,
i.e.\ prescription for which metric tensor is to be used in propagator
numerators and polarization sums. For the definition of ``internal''
and ``external'' see text.\label{tab:RSs}}
\end{table}

Note that the version of dimensional reduction denoted by \DR\ e.g.\
in Refs.~\cite{Kunszt:1993sd,Catani:1996pk} is equivalent to
\FDH\  at the one-loop  level (see e.g.\
Refs.~\cite{Kunszt:1993sd,Bern:1996je}).\footnote{%
In Ref.\ \cite{Bern:2002zk} a two-loop definition of the \FDH\ scheme has been
given. In what follows we will only use the one-loop definition and
the one-loop equivalence of \FDH\ and \DR.
}
 The infrared properties of
the three schemes \CDR, \HV, \DR\ (or equivalently \FDH) have been
studied and compared in Ref.~\cite{Catani:1996pk} and found to be
consistent with factorization.

An apparent inconsistency between dimensional reduction and
factorization has been identified in
Refs.~\cite{Beenakker:1988bq,Smith:2004ck}, but in these references
the version \DRED\ has been used. In Ref.~\cite{Signer:2005iu} it was
found that factorization holds as expected in \DRED\ if external
quasi-4-dimensional gluons are decomposed into $D$ dimensional gauge
fields and $(4-D)$ dimensional ``$\eps$-scalars'', which are treated
as separate partons. Technically, this decomposition amounts to
replacing
\begin{align}
\ASq{\ldots g\ldots}{}{\mDRED}&=
\ASq{\ldots \ghat\ldots}{}{\mDRED}+
\ASq{\ldots \gtilde\ldots}{}{\mDRED}
\nonumber\\
&=
\sum_{\gth\in\{\ghat,\gtilde\}}\ASq{\ldots \gth\ldots}{}{\mDRED}
,
\label{DREDdecomposition}
\end{align}
for squared matrix elements, where the different gluon types $g$,
$\ghat$, $\gtilde$ are denoted by the same symbols as the associated
metric tensors. The algebraic expressions for the partonic processes
involving $g$, $\ghat$, or $\gtilde$ are defined by the values of the
corresponding gluon polarization sums. These read
\begin{subequations}
\label{polsumsgphi}
\begin{align}
g:\quad&
\sum_{\rm pols}\epsilon^\mu\epsilon^\nu{}^* \to
-g^{\mu\nu}+\frac{n^\mu k^\nu+k^\mu n^\nu}{(nk)}
-\frac{n^2 k^\mu k^\nu}{(nk)^2},
\\
\ghat:\quad&
\sum_{\rm pols}\epsilon^\mu\epsilon^\nu{}^* \to
-\ghat^{\mu\nu}+\frac{n^\mu k^\nu+k^\mu n^\nu}{(nk)}
-\frac{n^2 k^\mu k^\nu}{(nk)^2},
\\
\gtilde:\quad&
\sum_{\rm pols}\epsilon^\mu\epsilon^\nu{}^* \to
-\gtilde^{\mu\nu}.
\end{align}
\end{subequations}
\Eqn{DREDdecomposition} follows trivially from
\Eqn{gghatgtilde}.

The decomposition of gluons into their $D$-dimensional and
$\eps$-scalar part in \DRED\ is also relevant for the renormalization
of UV divergences. In order to make Green functions with external
$\eps$-scalars finite, the renormalization constants for $\eps$-scalar
couplings in general have to be different from the corresponding gluon
couplings. For example, in pure QCD, the couplings $\alpha_s$ and
$\alpha_e$ for the quark--antiquark--gluon and the
quark--antiquark--$\eps$-scalar vertices receive different
counterterms $\delta \alpha_s\ne\delta\alpha_e$ even if
$\alpha_s=\alpha_e$ at tree level
\cite{Jack:1993ws,SteinhauserKantPapers}.

\subsection{Splittings in the four schemes\label{sec:splittings}}

An essential part of the \RS\ dependence of NLO contributions is
related to the \RS\ dependence of the splittings $i\to jk$ of one
parton $i$ into two collinear partons $j$, $k$. The \RS\ dependence of
real corrections is related to the splitting functions $P^{\mRS}_{i\to jk}$;
the \RS\ dependence of virtual corrections is related to constants
$\gamma_\mRS(i)$ \cite{Kunszt:1993sd}, which in turn can be derived from
the $P^\mRS_{i\to jk}$ via unitarity \cite{Catani:1996pk}. In this section
we explain the \RS\ dependence of the splitting functions and
correspondingly of the $\gamma_\mRS(i)$. The full results can be found in
Appendix~\ref{app:split}. 
\begin{figure}[tb]
\begin{picture}(400,100)
\SetOffset(20,20)
\Text(30,-20)[l]{\CDR}
\Text(25,28)[lb]{$\ghat$}
\Text(65,55)[lb]{$\ghat$}
\Text(74,41)[lb]{$\ghat$}
\Gluon(15,15)(44.6,30){2}{4}
\Gluon(44.6,30)(65.9,51.2){2}{4}
\Gluon(44.6,30)(73.6,37.8){2}{4}
\Line(10,10)(37.2,-2.7)
\Line(10,10)(39.9,7.4)
\Line(10,10)(29.3,-13.)
\GOval(10,10)(10,10)(0){0.8}
\SetOffset(120,20)
\Text(30,-20)[l]{\HV}
\Text(25,28)[lb]{$\bar{g}$}
\Text(65,55)[lb]{$\ghat$}
\Text(74,41)[lb]{$\ghat$}
\Gluon(15,15)(44.6,30){2}{4}
\Gluon(44.6,30)(65.9,51.2){2}{4}
\Gluon(44.6,30)(73.6,37.8){2}{4}
\Line(10,10)(37.2,-2.7)
\Line(10,10)(39.9,7.4)
\Line(10,10)(29.3,-13.)
\GOval(10,10)(10,10)(0){0.8}
\SetOffset(220,20)
\Text(30,-20)[l]{\FDH}
\Text(25,28)[lb]{$\bar g$}
\Text(65,55)[lb]{$g$}
\Text(74,41)[lb]{$g$}
\Gluon(15,15)(44.6,30){2}{4}
\Gluon(44.6,30)(65.9,51.2){2}{4}
\Gluon(44.6,30)(73.6,37.8){2}{4}
\Line(10,10)(37.2,-2.7)
\Line(10,10)(39.9,7.4)
\Line(10,10)(29.3,-13.)
\GOval(10,10)(10,10)(0){0.8}
\SetOffset(320,20)
\Text(30,-20)[l]{\DRED}
\Text(25,28)[lb]{$g$}
\Text(65,55)[lb]{$g$}
\Text(74,41)[lb]{$g$}
\Gluon(15,15)(44.6,30){2}{4}
\Gluon(44.6,30)(65.9,51.2){2}{4}
\Gluon(44.6,30)(73.6,37.8){2}{4}
\Line(10,10)(37.2,-2.7)
\Line(10,10)(39.9,7.4)
\Line(10,10)(29.3,-13.)
\GOval(10,10)(10,10)(0){0.8}
\end{picture}
\ccaption{}{ Gluon splitting into two collinear
  gluons in the four schemes, indicating the appropriate treatment of
  each gluon.\label{fig:splitggg} }
\end{figure}
Figure~\ref{fig:splitggg} shows the most interesting case of a gluon
splitting into two collinear gluons. According to the definition given
above, the two collinear gluons $j$ and $k$ are treated as
``internal'', and the virtual gluon $i$ as ``external''. The
appropriate treatment of the gluons in the four \RS\ can be read off
from Table~\ref{tab:RSs} and is displayed in the figure. Two simple
observations allow an easy comparison of the four cases.

First, the projection of a $D$-dimensional onto a strictly
4-dimensional parent gluon does not change the structure of the result
of the splitting functions. And second, the result in \DRED\ should be
decomposed according to \Eqn{DREDdecomposition} into four splittings
$\ghat\to\ghat\ghat$, $\ghat\to\gtilde\gtilde$,
$\gtilde\to\ghat\gtilde$, $\gtilde\to\gtilde\ghat$.\footnote{%
Splittings involving an odd number of $\gtilde$ vanish.
}
 Then the result in
\CDR\ is identical to the \DRED~result for $\ghat\to\ghat\ghat$, and
all scheme differences can be explained in the following way:
\begin{itemize}
\item The splitting $g\to gg$ is identical in the  \CDR\ and \HV\
  schemes. Formally, this is expressed in the equality
\begin{equation}
P^{<\, \mCDR}_{g^*\to g g}(z) = P^{<\, \mHV}_{g^*\to g g}(z)
 = P^{<\, \mDRED}_{\ghat^*\to \ghat \ghat} (z)
\end{equation}
for the splitting functions defined for $z<1$.
\item In the \FDH\ scheme the outgoing gluons are treated as
  quasi-4-dimensional. The resulting additional term can be
  interpreted as being due to the splitting
  $\ghat\to\gtilde\gtilde$ as already discussed in
  Ref.~\cite{Catani:1996pk}. Hence, 
\begin{equation}
P^{<\, \mFDH}_{g^*\to g g} (z) = P^{<\, \mDRED}_{\ghat^*\to \ghat \ghat} (z)
+ P^{<\, \mDRED}_{\ghat^*\to \gtilde \gtilde} (z) .
\end{equation}
\item In the \DRED\ scheme the parent gluon is also treated as
  quasi-4-dimensional, and therefore the two additional splittings
  $\gtilde\to\ghat\gtilde$ and $\gtilde\to\gtilde\ghat$ are
  possible. In the spirit of our discussion around
  \Eqn{DREDdecomposition} we do not combine the \DRED\ splitting
  functions into a single one.
\end{itemize}
The splitting functions involving quarks are related in a similar
way. Via unitarity, the \RS\ dependence of the constants $\gamma(i)$
follows from the splitting functions \cite{Catani:1996pk} and can thus
be explained in an analogous way:
\begin{itemize}
\item The $\gamma(i)$ in \CDR\ and \HV\ are the same,
\begin{equation}
\gamma_\mCDR(i) = \gamma_\mHV(i)\quad\mbox{for }i\in\{g,q\}.
\end{equation}
\item The additional terms in the \FDH\ scheme are due to the
  splittings $\ghat\to\gtilde\gtilde$ and $q\to q\gtilde$:
\begin{align}
\gamma_\mFDH(g)&=\gamma_\mHV(g)-
\int_0^1 dz\, z P^{<\, \mDRED}_{\ghat\to \gtilde\gtilde}(z) 
,
\\
\gamma_\mFDH(q)&=\gamma_\mHV(q)-\int_0^1 dz\, z
\left[P^{<\, \mDRED}_{q\to q \gtilde}(z)+
P^{<\, \mDRED}_{q\to  \gtilde q}(z)\right]
.
\end{align}
\item In \DRED\ one has to distinguish $\gamma$ constants
  for $\ghat$, $\gtilde$, and $q$. The ones corresponding
  to $\ghat$ and $q$ are the same as the ones in the \FDH\ scheme; the
  one for $\gtilde$ is related to the additional splittings 
  $\gtilde\to\ghat\gtilde$ and $\gtilde\to\gtilde\ghat$:
\begin{align}
\gamma_\mDRED(\ghat)&=\gamma_\mFDH(g),\\
\gamma_\mDRED(q)&=\gamma_\mFDH(q),\\
\gamma_\mDRED(\gtilde)&= -\int_0^1 dz\, z\,
\frac{(1-z)}{(1-z)_+}\left[P^{<\, \mDRED}_{\gtilde\to \gtilde \ghat}(z) 
\right.
\nonumber\\&\qquad\qquad
 \left.
   + P^{<\, \mDRED}_{\gtilde\to\ghat \gtilde}(z)
   + 2 N_F\, P^{<\, \mDRED}_{\gtilde\to q\bar{q}}(z)\right]
\end{align}
\end{itemize}
These relations form the basis for understanding the \RS\ dependence
of NLO contributions and in particular the difference between \DRED\
and the other schemes. In the subsequent sections we will see that
additional \RS\ dependences arise from the crossing of the splitting
functions to initial-state parton splitting and from the \RS\
dependence of the LO matrix element.


\section{CDR, HV, FDH}\label{sec:RS*}

Our starting point is the decomposition, \Eqn{sigmahat}, of the
hard partonic cross section, and we are mainly interested in the
\RS\ dependence of the separate terms contributing to $d\hat\sigma$.
In this section we will restrict ourselves to the well-known cases of
\CDR, \HV\ and \FDH. The \RS~ dependence of quantities will be
indicated by a subscript $\mRS*$, the star reminding us that we
consider \CDR, \HV\ and \FDH, but not (yet) \DRED.

\subsection{Born term} \label{sec:born}

We consider the partonic process
\begin{equation}
a_1(p_1)\,  a_2(p_2) \to  a_3(p_3)\ldots a_n(p_n)\  ,
\label{process}
\end{equation}
where $a_i$ and $p_i$ denote the flavour and the momentum of parton
$i$ respectively. The $l$-loop correction to the \RS\ dependent squared
matrix element for the process given in \Eqn{process} is denoted by
${\cal M}^{(l)}_{\mRS*}(a_1(p_1), a_2(p_2); a_3(p_3)\ldots a_n(p_n))$
or by ${\cal M}^{(l)}_{\mRS*}(a_1\ldots a_n)$ for short. For the cross
section we need the averaged squared matrix elements
\begin{equation}
\langle {\cal M}^{(l)}_{\mRS*}(a_1,a_2;\ldots a_n) \rangle = 
\frac{1}{2\, s_{12}} \frac{1}{\omega_{\mRS*}(a_1)\, \omega_{\mRS*}(a_2)}
\, {\cal M}^{(l)}_{\mRS*}(a_1,a_2;\ldots a_n) ,
\label{MavDef}
\end{equation}
where $\omega_{\mRS*}(a_i)$ denotes the \RS\ dependent number of
degrees of freedom of a parton with flavour $a_i$ and $s_{12} \equiv
2\, (p_1\cdot p_2)$ in the case of massless incoming partons.

The Born cross section is obtained by integrating the squared and
averaged tree-level matrix element over the $(n-2)$ parton phase space
$d\Phi_{n-2}(p_1,p_2;p_3\ldots p_n)$ multiplied by a measurement
function for an infrared-safe quantity and a symmetry factor. The
latter two are always implicitly understood in our notation and we
simply write
\begin{equation}
d\sigma^{(0)}_{\mRS*}(a_1\ldots a_n) =  
\int d\Phi_{n-2}(p_1\ldots p_n)\, 
\langle {\cal M}^{(0)}_{\mRS*}(a_1\ldots a_n) \rangle .
\label{SigmaNill}
\end{equation}
The \RS~dependence in \Eqn{SigmaNill} is due to ${\cal O}(\eps)$ terms
in ${\cal M}^{(0)}_{\mRS*}$. Since we consider an infrared-finite
quantity, the phase-space integration does not introduce any
poles. Therefore, we can take the limit $\eps\to 0$ and
\begin{equation}
d\sigma^{\rm Born}(a_1\ldots a_n) \equiv 
\left[d\sigma^{(0)}_{\mRS*}(a_1\ldots a_n) \right]_{D\to 4}
\label{SigmaBorn}
\end{equation} 
is \RS~independent, as indicated by the absence of the subscript
$\mRS*$.

\subsection{Virtual corrections} \label{sec:virtual}

For the virtual corrections we need ${\cal M}^{(1)}_{\mRS*}$, the
interference terms of the one-loop amplitude and the tree-level
amplitude.  The structure of the singular terms of ${\cal
M}^{(1)}_{\mRS*}$ is well known~\cite{Kunszt:1994mc,Catani:2000ef}.
For the fully renormalized matrix element, it is given by
\begin{eqnarray}
{\cal M}^{(1)}_{\mRS*}(a_1\ldots a_n) &=& 
\frac{\alpha_s}{2\pi} c_\Gamma   \Bigg[
 {\cal M}^{(0)}_{\mRS*}(a_1\ldots a_n) \left(
    - \frac{1}{\epsilon} \sum_{i} \gamma_{\mRS*}(a_i) \right)
\label{Mvirtual*}
\\ \nonumber
&& \hspace*{-1cm}
 +\ \sum_{i, j} {\cal V}(i,j)\, 
     {\cal M}^{ij}_{\mRS*}(a_1\ldots a_n)
   + {\cal M}_\mNS^{(1)}(a_1\ldots a_n) \Bigg]  ,
\end{eqnarray}
where the sums $i$ and $j$ are over all initial or final state partons and we
introduced
\begin{equation}
c_\Gamma \equiv (4\pi)^{\eps}\, 
\frac{\Gamma(1+\eps)\Gamma^2(1-\eps)}{\Gamma(1-2\eps)}  =
\left(\frac{e^{\gamma_E}}{4\pi}\right)^{-\eps} 
\left(1-\frac{\eps^2\, \pi^2}{12} + {\cal O}(\eps^3)\right) \, .
\label{def:cGamma}
\end{equation}
The soft and collinear poles are contained in the terms proportional
to ${\cal M}^{(0)}_{\mRS*}$ and ${\cal M}^{ij}_{\mRS*}$. The latter
are the colour-linked Born squared matrix elements introduced in
Ref.~\cite{Kunszt:1992tn} and correspond to the square of the
colour-correlated tree amplitudes with a ${\bf T}_i\cdot {\bf T}_j$
insertion, used in Ref.\cite{Catani:1996vz}. If particles $i$ and $j$
are massless, we have
\begin{equation}
{\cal V}(i,j) = -\frac{1}{2 \eps^2}\, 
{\rm Re}\! \left(-\frac{s_{ij}}{\mu^2}\right)^{-\eps} 
\label{Vij_massless}
\end{equation}
and \Eqn{Mvirtual*} reduces to the well-known expression for the
singularities of one-loop QCD amplitudes~\cite{Kunszt:1994mc}. If one
or both of the particles $i,\ j$ are massive, these expressions have
to be generalized~\cite{Catani:2000ef}, but the structure of the
singularities remains as in \Eqn{Mvirtual*}. 

The \RS\ dependence of \Eqn{Mvirtual*} is contained in the constants
$\gamma_{\mRS*}(a_i)/\eps$, as well as in ${\cal M}^{(0)}_{\mRS*}$ and
${\cal M}^{ij}_{\mRS*}$. The remaining term, ${\cal
M}_\mNS^{(1)}(a_1\ldots a_n)$, is in general very complicated, but is
finite and, after taking the limit $D\to 4$, \RS\ independent.

The essential, non-trivial part of the \RS\ dependence is due to the
$\gamma_{\mRS*}(a_i)/\eps$ terms. These are closely related to
collinear singularities due to self-energy insertions on external legs
and depend on the flavour $a_i$ of leg $i$. 
The \RS~dependence of $\gamma_{\mRS*}$ has been given in
Ref.~\cite{Kunszt:1993sd}. Via unitarity it is related to parton 
splittings $a_i\to$anything \cite{Catani:1996pk}, as summarized in
Section~\ref{sec:splittings}. 

In the present paper we determine $\gamma_{\mRS*}$ by insisting that
the sum rules hold in all \RS\ to all orders in $\eps$. This is a
slightly different approach compared to Ref.~\cite{Kunszt:1993sd} and
simply amounts to a shift of finite terms between ${\cal
M}_\mNS^{(1)}$ and the $\gamma_{\mRS*}$ terms in
\Eqn{Mvirtual*}. Neglecting ${\cal O}(\eps^2)$ terms, we find
\begin{align}
\gamma_\mCDR(g) &= \gamma_\mHV(g) = 
    \frac{\beta_0}{2} + \epsilon\, \frac{T_F N_F}{3}; &\!\! 
    \gamma_\mCDR(q)& = \gamma_\mHV(q) = \frac{3 C_F}{2} + \epsilon\,
    \frac{C_F}{2}
\nonumber\\
\gamma_\mFDH(g) &= 
    \frac{\beta_0}{2} + \epsilon\, \frac{2 T_F N_F-N_c}{6}; 
&\!\! \gamma_\mFDH(q)& = \frac{3 C_F}{2}
\label{gammaALL}
\end{align}
with $\beta_0 = (11 N_c - 4\, T_F\, N_F)/3$
and $T_F=1/2$. For heavy quarks the
result is \RS~independent~\cite{Catani:2000ef} and we have $\gamma(Q)
= C_F$.

The final virtual corrections $d\sigma_{\mRS*}^{\rm virt}$ are obtained as
\begin{equation}
d\sigma^{\rm virt}_{\mRS*}(a_1\ldots a_n) =  
\int d\Phi_{n-2}(p_1\ldots p_n)\, 
\langle {\cal M}^{(1)}_{\mRS*}(a_1\ldots a_n) \rangle .
\label{SigmaVirt}
\end{equation}
Since this phase space integration does not give rise to any pole in
$\eps$, taking the limit $D\to 4$ in ${\cal M}_\mNS^{(1)}$ is
justified.

\subsection{Real corrections} \label{sec:real}

For the real corrections $d\sigma^{\rm real}_{\mRS*}$ to the partonic
process $a_1\, a_2 \to a_3\ldots a_n$ we have to consider
contributions 
from all $2\to (n-1)$ processes that are obtained by a
split of any of the outgoing partons. We have to evaluate the
corresponding squared matrix elements ${\cal
M}^{(0)}_\mRS(a_1,a_2;\bar a_3 \ldots \bar a_{n+1})$ and integrate
them over the $(n-1)$ parton phase space $d\Phi_{n-1}(p_1,p_2;p_3\ldots
p_{n+1})$
\begin{equation}
d\sigma^{\rm real}_{\mRS*} = \sum_{\bar a_i} \int
 d\Phi_{n-1}(p_1,p_2;p_3\ldots p_{n+1}) \langle {\cal
 M}^{(0)}_{\mRS*}(a_1,a_2;\bar a_3 \ldots \bar a_{n+1}) \rangle .
\label{RealFull}
\end{equation}
In \Eqn{RealFull} we denote by $\bar a_i$, $i\in\{3\ldots n+1\}$ the
flavour of the outgoing partons, and as indicated by $\sum_{\bar a_i}$,
we have to sum over all relevant processes.

As is well known, the matrix elements can develop singularities in
regions of the phase space where a parton becomes soft or two partons
become collinear. The integration over $d\Phi_{n-1}(p_1\ldots
p_{n+1})$ in this region then results in $1/\eps^2$ and $1/\eps$
poles. Thus the \RS~dependence of the matrix elements which manifests
itself in the ${\cal O}(\eps)$ terms of ${\cal M}^{(0)}_\mRS$ results
in differences in the ${\cal O}(1/\eps)$ and in the finite terms of
the real corrections.\footnote{Note that in the \HV\ and \FDH\ schemes
real  soft and/or collinear gluons have to be treated as ``internal'',
i.e.\ in the same way as gluons in a closed loop but differently from
observed, ``external'' gluons. This is the only source of the \RS\
dependence of the real corrections in these schemes.}

In order to deal with the phase-space integration at NLO one often
uses either phase-space slicing \cite{Giele:1991vf} or subtraction
\cite{Ellis:1980wv,Kunszt:1992tn}, and 
several general procedures have been developed~\cite{Frixione:1995ms,
Catani:1996vz, more}. For our purposes it is sufficient to know that
they all rely on the same main points.  Using the results given below
it will be obvious how any of these procedures can be applied in the
context of \DRED. 

The first point is that in any of the singular regions the matrix
elements take a simple form and can be written as a factor containing
the kinematic singularity times a reduced (colour-linked) tree-level
matrix element, associated with a $2\to (n-2)$ process. Secondly,
the phase space is factorized according to $d\Phi_{n-1}(p_1\ldots
p_{n+1}) = d\Phi_{n-2}(p'_1\ldots p'_n)\, d\Phi_{\rm rad}$. The factor
with the kinematic singularity is integrated analytically over
$d\Phi_{\rm rad}$, producing the poles in analytic form. These poles
will be multiplied by the reduced (colour-linked) matrix element and
are to be integrated over a slightly modified $(n-2)$ parton phase space
$d\Phi_{n-2}(p'_1\ldots p'_n)$. It is therefore not surprising that
the real corrections have a similar structure as the virtual
corrections, \Eqns{Mvirtual*}{SigmaVirt}. We will now look at all
three potentially singular regions in turn.

\medskip
\emph{Soft Region:} In the limit where gluon $g_k$ (or another
massless gauge boson) becomes soft we have
\begin{equation}
{\cal M}^{(0)}_{\mRS*}(a_1,a_2;\ldots g_{k}(p_{k})\ldots \bar a_{n+1})
\stackrel{p_{k}\to 0}{=}
g^2_s\,  \sum_{i,j}
\frac{s_{ij}}{s_{ik} s_{jk}}\, 
{\cal M}^{ij}_{\mRS*}(a_1\ldots a_n) ,
\label{SoftLim}
\end{equation}
where we introduced $g^2_s = 4\pi\, \alpha_s$.  It is understood that
the set of partons $\{a_3\ldots a_n\}$ is equal to the set
  $\{\bar{a}_3\ldots g_k\ldots\bar{a}_{n+1}\}$ with $g_k$
  removed. Similar comments apply to the analogous equations below.
The phase space integration of \Eqn{SoftLim} leads to $\int
d\Phi_{\rm rad}\, s_{ij}/(s_{ik} s_{jk})$ and results in poles that cancel
the corresponding poles in ${\cal V}(i,j)$. Due to the measurement
function implicitly included in $d\Phi_{n-2}$, the remaining
integration does not result in any singularities. The scheme
dependence enters only through ${\cal M}^{ij}_{\mRS*}$ and after
summation over all real processes trivially cancels between the real
and virtual corrections.

\medskip
\emph{Final-State Collinear Region:} In the limit where two outgoing
partons $\bar a_k$ and $\bar a_l$ become collinear we have
\begin{eqnarray}
\lefteqn{{\cal M}^{(0)}_{\mRS*}
    (a_1,a_2;\ldots \bar a_l(p_l)\ldots \bar a_k(p_k)\ldots \bar a_{n+1}) 
\stackrel{p_k\| p_l}{=} } 
\label{CollLim} \\
&& \frac{2\, g_s^2}{s_{kl}}
P_{{(kl)}^*\to k l}^{<\, \mRS*}(z)\, 
{\cal M}^{(0)}_{\mRS*}(a_1,a_2;\ldots a_{(kl)}(p_k+p_l)\ldots a_{n}) .
\nonumber
\end{eqnarray}
As detailed in Appendix~\ref{app:split}, $P_{{(kl)}^*\to k l}^{<\
\mRS*}$ is the \RS\ dependent splitting function defined for $z<1$ with
$p_k\to z (p_k+p_l)$ and $p_l \to (1-z)(p_k+p_l)$. In \Eqn{CollLim}
the flavours $\bar a_k$ and $\bar a_l$ are fixed. This uniquely
determines the flavour of the parent parton $a_{(kl)}$. It is
understood that if the split is flavour forbidden we set
$P_{{(kl)}^*\to k l}^{<\ \mRS*} = 0$. To avoid a proliferation of
subscripts, we denote the flavour of the partons in the splitting
functions simply by $(kl)$ etc. rather than $a_{(kl)}$. The parent
parton is slightly off shell as indicated by the notation $(kl)^*$.

Contrary to the soft limit, in the collinear limit there are two
sources of \RS~dependence. Apart from the trivial dependence through
${\cal M}^{(0)}_{\mRS*}(a_1\ldots a_n)$, the prefactor $P_{{(kl)}^*\to
k l}^{<\, \mRS*}$ is also \RS~dependent.  Its \RS~dependence can be
found in Section~\ref{sec:splittings} and in
Appendix~\ref{app:split}. Since the $z$ dependence in \Eqn{CollLim} is
entirely in the prefactor, the integration $\int d\Phi_{\rm rad}\,
P_{{(kl)}^*\to k l}^{<\, \mRS*}/s_{kl}$ can be performed
separately. The terms related to the collinear singularities due to
the splitting of parton $a_i = a_{(kl)}$ schematically can be written
as
\begin{eqnarray}
\label{outColl}
d\sigma^{{\rm real},i}_{\mRS*}(a_1\ldots a_i\ldots a_n)  &=&
  - \frac{\alpha_s}{2\pi} \frac{c_\Gamma}{\eps}\,
   d\sigma^{(0)}_{\mRS*}(a_1\ldots a_i\ldots a_n)
\\
\nonumber
&\times&
\sum_{a_k}
\int dz\, \Theta\left(z-\frac{1}{2}\right) 
\frac{(1-z)}{(1-z)_+} P_{i^*\to kl}^{<\, \mRS*}(z)\, ,
\end{eqnarray}
where the sum over all possible splittings, $\sum_{a_k}$, is due to
the sum over the relevant real processes, \Eqn{RealFull}. Note that in
the sum $a_k\in\{g,q,\bar q\}$ a sum over the $N_F$ massless quark
flavours is implicitly understood.   After this sum the
integrand is symmetric with respect to $z\leftrightarrow 1-z$. Hence
the integration can be restricted to the region $z>1/2$ and the
potential singularity at $z=1$ is regularized with the usual
$+$prescription. The $z$ integration in \Eqn{outColl} results in a
\RS~dependence of the singular pre\-factor that multiplies the Born
term. In fact, the factor in the second line of \Eqn{outColl} is equal
to $(-\gamma_{\mRS*}(a_i))$. Therefore, after summing up the
contributions of all final 
state partons, $i\in\{3\ldots n\}$ these terms precisely
cancel the singularity and the \RS\ dependence of those virtual terms
displayed in the first line of \Eqn{Mvirtual*} that are associated
with outgoing partons, $i \ge 3$ \cite{Catani:1996pk}.

\begin{figure}[t]
\begin{center}
\begin{picture}(250,70)(0,-10)
\SetOffset(20,0)
\Text(-10,45)[l]{$\gamma_{\mRS*}(g)\big|_{N_c}$}
\Gluon(15,15)(44.6,30){2}{4}
\Gluon(44.6,30)(65.9,51.2){2}{4}
\Gluon(44.6,30)(73.6,37.8){2}{4}
\Line(10,10)(37.2,-2.7)
\Line(10,10)(39.9,7.4)
\Line(10,10)(29.3,-13.)
\GOval(10,10)(10,10)(0){0.8}
\SetOffset(170,0)
\Text(-40,15)[r]{$+$}
\Text(-10,45)[l]{$\gamma_{\mRS*}(g)\big|_{N_F}$}
\Gluon(15,15)(44.6,30){2}{4}
\ArrowLine(44.6,30)(65.9,51.2)
\ArrowLine(73.6,37.8)(44.6,30)
\Line(10,10)(37.2,-2.7)
\Line(10,10)(39.9,7.4)
\Line(10,10)(29.3,-13.)
\GOval(10,10)(10,10)(0){0.8}
\end{picture}
\end{center}
\ccaption{}{Illustration of \Eqn{outColl} for the case   where $a_i$
  is an
outgoing splitting gluon. The sum over all relevant real processes,
$\sum_{a_k}$, gives rise to two contributions. The one on the left (right)
results in the $N_c$ ($N_F$) part of $\gamma_{\mRS*}(g)$.
\label{fig:outA}}
\end{figure}

\medskip 
\emph{Initial-State Collinear Region:} Finally we turn to the case of
an outgoing parton $\bar a_k$ becoming collinear to the incoming
parton $a_1$ (or $a_2$). There are some important differences with
respect to \Eqn{CollLim}. To start with, the collinear limit has to be
written with the spin/colour summed/averaged matrix elements $\langle
{\cal M}^{(0)}_{\mRS*} \rangle$. With $p_k \to (1-z) p_1$ the
collinear limit is given by
\begin{eqnarray}
\lefteqn{\langle{\cal M}^{(0)}_{\mRS*}(a_1(p_1), a_2;\ldots \bar
  a_{k}(p_k) \ldots\bar a_{n+1})  
\rangle \stackrel{p_k\| p_1}{=} } 
\label{InCollLim} \\
&& \frac{2\, g_s^2}{s_{1k}} 
P_{1\to {(1k)}^*k}^{<\, \mRS*}(z)\,
\langle{\cal M}^{(0)}_{\mRS*}(a_{(1k)}(z\, p_1), a_2;\ldots a_{n})
\rangle .
\nonumber
\end{eqnarray}
Contrary to \Eqn{CollLim}, the $z$ dependence in \Eqn{InCollLim} is
not restricted to the prefactor and, therefore, the $z$-integration
results in a more complicated structure.
In \Eqn{InCollLim} the flavours $a_1$ and $\bar{a}_k$ are fixed and
uniquely determine the flavour of parton $a_{(1k)}$, which is slightly
off shell. The splitting functions of \Eqns{CollLim}{InCollLim}, with
initial/final-state off-shell parton, are related by a crossing
relation
\begin{equation}
P_{l\to (lk)^*k}^{<\, \mRS*}(z) = (-1)^{\#f+1} 
\frac{\omega_\mRS(a_{(lk)})}{\omega_{\mRS*}(a_{l})}\, 
z\, P_{(lk)^*\to l k}^{<\, \mRS*}\left(\frac{1}{z}\right) ,
\label{Pcrossing}
\end{equation}
where $\#f$ denotes the number of crossed fermions. We remark that the
well-known crossing symmetry $P_{l\to (lk)^* k}^{<\, \mCDR} =
P_{l^*\to (lk) k}^{<\, \mCDR}$ does not hold in all \RS, and
in general we have
\begin{equation}
P_{l\to (lk)^*k}^{<\, \mRS*}(z) = 
P_{l^*\to (lk) k}^{<\, \mRS*}(z) + 
\Delta_{l\to (lk) k}^{\mRS*}(z)
\label{CrossingViolation}
\end{equation}
with $\Delta^{\mFDH}\neq 0$ and $\Delta^{\mHV}\neq 0$. The explicit
form of $\Delta^{\mFDH}$ and $\Delta^{\mHV}$ can easily be found using
\Eqn{Pcrossing} and the results in Appendix~\ref{app:split}. However,
as will be discussed in Section~\ref{sec:incoll}, the distinction made
in \Eqn{CrossingViolation} is ultimately not required.

Using \Eqns{CrossingViolation}{FullP} to express the collinear limit,
\Eqn{InCollLim}, in terms of the full splitting functions $P_{1\to
(1k) k}^{\mRS*}$ and summing over all relevant real processes, we can
write the initial state collinear term for parton 1 schematically as
\begin{eqnarray}
\label{inColl}
d\sigma^{{\rm real},1}_{\mRS*}(a_1\ldots a_n) &=& 
   \frac{\alpha_s}{2\pi} \frac{c_\Gamma}{\eps}
\bigg[\gamma_{\mRS*}(a_1)\, 
d\sigma^{(0)}_{\mRS*}(a_1(p_1),a_2;\ldots a_n)  
\\ \nonumber
&& \hspace*{-2cm} -\  \sum_{a_k} \int dz\
\left( P_{1\to (1k) k}^{\mRS*}
       + \Delta_{1\to (1k) k}^{\mRS*} \right)
d\sigma^{(0)}_{\mRS*}(a_{(1k)}(z\, p_1),a_2;\ldots a_n) \bigg]  .
\end{eqnarray}
The first term on the r.h.s. of \Eqn{inColl} is due to the
$\delta(1-z)$ term present in \Eqn{FullP}. Together with the
corresponding term for the second incoming parton, $d\sigma^{{\rm
real},2}_{\mRS*}$, this results in a term that precisely cancels the
singularity and the \RS\ dependence of those virtual terms displayed
in the first line of \Eqn{Mvirtual*} that are associated with incoming
partons $i \le 2$. The remaining terms given in the second line on the
r.h.s. of \Eqn{inColl} are associated with collinear counterterms.

\begin{figure}[t]
\begin{center}
\begin{picture}(340,60)(0,0)
\SetOffset(-20,0)
\Gluon(65,15)(36.4,20.4){2}{4}
\Gluon(36.4,20.4)(15.0,20.4){2}{3}
\Gluon(36.4,20.4)(89.3,36.8){2}{8}
\Line(70,10)(97.2,-2.7)
\Line(70,10)(91.2,-11.2)
\Line(70,7)(31.4,-3.4)
\GOval(70,10)(10,10)(0){0.8}
\SetOffset(110,0)
\Text(-5,15)[r]{$+$}
\Gluon(36.4,20.4)(15.0,20.4){2}{3}
\ArrowLine(65,15)(36.4,20.4)
\ArrowLine(36.4,20.4)(89.3,36.8)
\Line(70,10)(97.2,-2.7)
\Line(70,10)(91.2,-11.2)
\Line(70,7)(31.4,-3.4)
\GOval(70,10)(10,10)(0){0.8}
\SetOffset(240,0)
\Text(-5,15)[r]{$+$}
\Gluon(36.4,20.4)(15.0,20.4){2}{3}
\ArrowLine(36.4,20.4)(65,15)
\ArrowLine(89.3,36.8)(36.4,20.4)
\Line(70,10)(97.2,-2.7)
\Line(70,10)(91.2,-11.2)
\Line(70,7)(31.4,-3.4)
\GOval(70,10)(10,10)(0){0.8}
\end{picture}
\end{center}
\ccaption{}{Illustration of \Eqn{inColl} for an incoming splitting
  gluon. The sum over all relevant real processes, $\sum_{a_k}$,
  gives rise to three terms. \label{fig:inA}}
\end{figure}

\subsection{Collinear counterterm} \label{sec:incoll}

In the sum of the  virtual and real   corrections, 
$d\sigma^{\rm virt}_{\mRS*} + d\sigma^{\rm real}_{\mRS*}$,
all singularities and \RS~dependences cancel,
apart from the terms given in the last line of \Eqn{inColl}.
These are cancelled by the collinear counterterm $d\sigma^{\rm
coll}_{\mRS*}$. While the divergent parts of the collinear
counterterms are completely determined, there is some freedom in how
to specify the finite parts of $d\sigma_{\mRS*}^{\rm coll}$. Any
specific choice of the finite parts of $d\sigma_{\mRS*}^{\rm coll}$ is
equivalent to the definition of a particular factorization scheme.  
Leaving the factorization scheme open, we can write
\begin{eqnarray}
\label{NonMScoll}
\lefteqn{d\sigma^{\rm coll}_{\mRS*,\mFS}(a_1,a_2;\ldots a_n) =
\frac{\alpha_s}{2\pi} \frac{c_\Gamma}{\eps} \sum_{a_k} \int dz} && 
\\ \nonumber
&\times& \bigg[ 
\left( P_{1\to i k}^{\mRS*}(z) 
       + \Delta_{1\to i k}^{\mRS*}(z) 
       + \epsilon X^{\mFS}_{1\to ik}(z) \right)
d\sigma^{(0)}_{\mRS*}(a_i(z\, p_1),a_2(p_2);\ldots a_n) 
\\ \nonumber
&& + \,  
\left( P_{2\to i k}^{\mRS*}(z) 
       + \Delta_{2\to i k}^{\mRS*}(z) 
       + \epsilon X^{\mFS}_{2\to ik}(z) \right)
d\sigma^{(0)}_{\mRS*}(a_1(p_1),a_i(z\, p_2);\ldots a_n) \bigg]\ ,
\end{eqnarray}
where the sum is over all possible splittings of the incoming partons,
and the index $i$ is defined in analogy to \Eqns{CollLim}{outColl}.
The $X^{\mFS}_{l\to ik}$ are the finite (i.e.\
$\epsilon$-independent), \RS\ independent terms which define the
factorization scheme.  The formulas in Sections~\ref{sec:virtual} and
\ref{sec:real} show that in this way the hard (subtracted) partonic cross
section
\begin{eqnarray}
\label{subNonMS}
\lefteqn{d\hat\sigma_{\mFS}(a_1\ldots a_n) = \Big[ 
d\sigma^{\rm Born}_{\mRS*} (a_1\ldots a_n)} &&
\\ \nonumber &&
+\ d\sigma^{\rm virt}_{\mRS*} (a_1\ldots a_n)
+ d\sigma^{\rm real}_{\mRS*}(a_1\ldots a_n)
+ d\sigma^{\rm coll}_{\mRS*,\mFS}(a_1\ldots a_n) \Big]_{D\to 4}
\end{eqnarray}
is finite and \RS~independent as indicated by the absence of the
subscript $\mRS$. In this approach the most natural factorization
scheme would correspond to setting all $X^{\mFS}=0$. In principle,
this particular factorization scheme is as good as any other, and it
could be used in practice in \Eqn{SigmaHad} in conjunction with parton
distribution functions $f_{a_i/H}$ determined in the same scheme. In
practice, however, parton distribution functions such as the standard
MRST or CTEQ sets \cite{MRST,CTEQ} are mainly available in the
\MSbar\ factorization scheme, which is different.

The \MSbar\ factorization scheme is defined by using \RS=\CDR\ and
replacing the square bracket  in \Eqn{NonMScoll} by
\begin{eqnarray}
\label{MScoll}
\bigg[
 \left[P_{1\to i k}^{\mCDR}(z)\right]_{D\to 4}\, 
d\sigma^{(0)}_\mCDR(a_{i}(z\, p_1),a_2;\ldots a_n) 
+\{ 1\leftrightarrow 2\} \, 
\bigg].
\end{eqnarray}
The minimal subtraction procedure corresponds to setting $\eps\to 0$
in the splitting functions $ P_{1\to i k}^{\mCDR}$. Thus, even if we
use \CDR\ (and even though $\Delta_{1\to i k}^{\mCDR} = 0$) the
\MSbar\ scheme does not correspond to $X^{\mFS}=0$ but to
\begin{equation}
\epsilon X^{\mMS}_{1\to ik}(z) =
-P_{1\to i k}^{\mMS\, \eps}(z)
,
\end{equation}
where $P_{1\to i k}^{\mMS\, \eps}(z) \equiv P_{1\to i k}^{\mCDR}(z) -
\left[ P_{1\to i k}^{\mCDR}(z) \right]_{D\to 4}$ denote the ${\cal
O}(\eps)$ terms of the splitting functions in \CDR.

In the evaluation of the real corrections, \Eqn{inColl}, as well as in
the collinear counterterm, \Eqn{NonMScoll}, we made the distinction
between $P^{<\, \mRS*}_{l^*\to ik}$ and $P^{<\, \mRS*}_{l\to i^*k}$,
the splitting functions appropriate for an outgoing and incoming split,
respectively.  In a general \RS, these two splitting functions differ
as indicated in \Eqn{CrossingViolation}, hence the presence of the
$\Delta$ terms in \Eqns{inColl}{NonMScoll}. While the expressions
given in \Eqns{inColl}{NonMScoll} are those that naturally arise in
the calculation, we note that the $\Delta$ terms cancel in the sum of
$d\sigma^{\rm real}_{\mRS*} + d\sigma^{\rm coll}_{\mRS*,\mFS}$. Thus
we can drop $\Delta^{\mRS*}_{1\to (1k)k}$ in \Eqn{inColl} (and the
corresponding term in $d\sigma^{{\rm real},2}_{\mRS*}$) if we also
drop $\Delta^{\mRS*}_{1\to ik}$ (and $\Delta^{\mRS*}_{2\to ik}$) in
\Eqn{NonMScoll}. Ultimately, the distinction between $P^{<\,
\mRS*}_{l^*\to ik}$ and $P^{<\, \mRS*}_{l\to i^*k}$ is not needed.

\section{Dimensional Reduction} \label{sec:DRED}

In this section we show how the structure described in
Section~\ref{sec:RS*} can be generalized to include \DRED. As
mentioned in the introduction, the key point is to split the gluon
into a $D$-dimensional gluon $\ghat$ and a $(4-D)$-dimensional
$\eps$-scalar $\gtilde$ by setting $g=\ghat+\gtilde$ and using
\Eqn{DREDdecomposition}.  Often it is sufficient to perform this split
for only one gluon as in \Eqn{DREDdecomposition}; in general, if a
process contains $\#g$ gluons, $g_{i_1} \ldots g_{i_{\#g}}$,
we can decompose the matrix element in \DRED\ into $2^{\#g}$ terms according to
\begin{equation}
\label{DREDsplitI}
{\cal M}^{}_\mDRED(\ldots g_{i_1} \ldots g_{i_{\#g}}\ldots) =
\hspace*{-0.2cm}
\sum_{\gth_{i_1}\in\{\ghat,\gtilde\}}\! \ldots\!
\sum_{\gth_{i_{\#g}}\in\{\ghat,\gtilde\}}\! 
{\cal M}^{}_\mDRED(\ldots \gth_{i_1}\ldots \gth_{i_{\#g}}\ldots ).
\end{equation}
We consider the two partons $\ghat$ and $\gtilde$ to be two different
partons and consequently regard the r.h.s of \Eqn{DREDsplitI} as a sum
over the squared matrix elements of $2^{\#g}$ different processes. To
bring our notation in line with the previous section, we will write
\Eqn{DREDdecomposition} and
\Eqn{DREDsplitI} as
\begin{align}
\label{DREDspliti}
{\cal M}^{}_\mDRED(a_1\ldots a_i \ldots a_n) &= 
\sum_{\ath_i}{\cal M}^{}_\mDRED(a_1\ldots \ath_i\ldots a_n) 
,
\\
\label{DREDsplit}
{\cal M}^{}_\mDRED(a_1\ldots a_n) &= 
\sum_{\{\ath\}}{\cal M}_\mDRED(\ath_1\ldots \ath_n) ,
\end{align}
respectively, where it is understood that if $a_i = g$ we sum over the
two terms $\ath_i\in\{\ghat, \gtilde\}$ whereas if $a_i = q$ there is
only one term in the sum $\ath_i\in\{q\}$. 

For the spin summed/averaged matrix elements the relation equivalent
to \Eqn{DREDsplit} reads
\begin{equation}
\label{DREDsplitAV}
\langle {\cal M}^{}_\mDRED(a_1,a_2;\ldots a_n) \rangle = 
\sum_{\{\ath\}} 
\frac{\omega_\mDRED(\ath_1)}{\omega_\mDRED(a_1)}\
\frac{\omega_\mDRED(\ath_2)}{\omega_\mDRED(a_2)}\
\langle {\cal M}^{}_\mDRED(\ath_1,\ath_2;\ldots \ath_n)  \rangle.
\end{equation}
The explicit expressions for $\omega_\mRS(a_i)$ are given in the
Appendix in \Eqn{omegas}.

We stress that while the split $g=\ghat+\gtilde$ is conceptually
simple, it seems to complicate practical computations. As we will see
in the later sections, however, in an explicit computation of a
physical process in \DRED\ it is only required at a very limited
number of steps.  In particular, it will turn out that no PDF for the
unphysical $\eps$-scalar $\gtilde$ will be required.

In the present section we will use the split to understand the
infrared structure of matrix elements in \DRED. In fact, it is
straightforward to see that Eqs.\ (\ref{Mvirtual*}), (\ref{CollLim}),
and (\ref{InCollLim}) for the collinear singularities of virtual and
real corrections hold in the same form in \DRED\ for the individual
processes with split partons, i.e.\ if we replace $\mRS*\to\mDRED$ and
$a_i\to\ath_i$ in these equations.  However, our main interest are the
infrared properties and \RS\ dependences of matrix elements for full
gluons; therefore we will carry out the sums over $\{\ath_i\}$
wherever possible.

\subsection{Born term} \label{sec:bornDRED}

The full tree-level matrix element in \DRED\ is equal to the one in
\FDH\ and \HV\ and can be obtained from the \CDR\ result simply by
setting $D\to 4$
\begin{eqnarray}
{\cal M}^{(0)}_\mDRED(a_1\ldots a_n) &=&
{\cal M}^{(0)}_\mHV(a_1\ldots a_n) =
{\cal M}^{(0)}_\mFDH(a_1\ldots a_n)
\label{Mtree}\\ 
&=&
\left[{\cal M}^{(0)}_\mCDR(a_1\ldots a_n)\right]_{D\to 4}   .
 \nonumber 
\end{eqnarray}
The Born cross section in \DRED\ can be obtained from \Eqn{Mtree}
and satisfies
\begin{equation}
d\sigma^{(0)}_\mDRED(a_1\ldots a_n) 
= d\sigma^{(0)}_\mHV(a_1\ldots a_n) = d\sigma^{(0)}_\mFDH(a_1\ldots
a_n) .
\end{equation}

\subsection{Virtual corrections} \label{sec:virtualDRED}

The structure of the virtual corrections in \DRED\ is analogous to
\Eqn{Mvirtual*} for each individual \DRED\ process with split partons
$\ath_i$. Hence, by summing over all processes as in \Eqn{DREDsplit}
we obtain 
\begin{eqnarray}
\label{Mvirtual}
{\cal M}^{(1)}_{\mDRED}(a_1\ldots a_n)
& = &
\sum_{\{ \ath \}}\, 
\frac{\alpha_s}{2\pi} c_\Gamma \Bigg[
 {\cal M}^{(0)}_{\mDRED}(\ath_1\ldots \ath_n) \left(
    - \frac{1}{\epsilon} \sum_{i} \gamma_{\mDRED}(\ath_i) \right)
\\ \nonumber
& + &
 \sum_{i, j} {\cal V}(i,j)\, 
     {\cal M}^{ij}_{\mDRED}(\ath_1\ldots \ath_n)
   + {\cal M}_\mNS^{(1)}(\ath_1\ldots \ath_n) \Bigg]  .
\end{eqnarray}
The non-trivial structure of \Eqn{Mvirtual} is most essential for the
$\gamma$ terms, since $\gamma_\mDRED(\ghat) \neq
\gamma_\mDRED(\gtilde)$. In fact, as discussed in
Section~\ref{sec:splittings} and Appendix~\ref{app:split}, the
$\gamma$ for $\ghat$ and $q$ match the ones in the \FDH\ scheme,
while the one for $\gtilde$ is new and different:
\begin{subequations}
\begin{align}
\gamma_\mDRED(\ghat)&=\gamma_\mFDH(g) =\frac{\beta_0}{2} +
    \epsilon\, \frac{2 T_F N_F-N_c}{6}
,\\
\gamma_\mDRED(q)&=\gamma_\mFDH(q)= \frac{3 C_F}{2},
\\
\gamma_\mDRED(\gtilde)&=2 N_c -T_F N_F.
\end{align}
\label{gammaDRED}
\end{subequations}

On the other hand, ${\cal V}(i,j)$ is not affected if a $\ghat$ is
replaced by a $\gtilde$, and the Born terms in the last line on the
r.h.s. of \Eqn{Mvirtual} can be combined in a straightforward way.
Hence one can immediately obtain the result for either
the fully or  the partially combined
process, ${\cal M}^{(1)}_{\mDRED}(a_1\ldots a_n)$ or ${\cal
  M}^{(1)}_{\mDRED}(\ath_1,\ath_2;a_3\ldots a_n)$, if 
desired. In particular, the \DRED\ result for the process involving
only full gluons satisfies
\begin{eqnarray}
{\cal M}^{(1)}_{\mDRED}(a_1\ldots a_n)
& = &
\frac{\alpha_s}{2\pi} c_\Gamma \Bigg[
\sum_i\sum_{\ath_i}\, 
 {\cal M}^{(0)}_{\mDRED}(a_1\ldots \ath_i\ldots a_n) \left(
    - \frac{1}{\epsilon} \gamma_{\mDRED}(\ath_i) \right)
 \nonumber
\\
& + &
 \sum_{i, j} {\cal V}(i,j)\, 
     {\cal M}^{ij}_{\mDRED}(a_1\ldots a_n)
   + {\cal M}_\mNS^{(1)}(a_1\ldots a_n) \Bigg]  ,
\label{MvirtualFull}
\end{eqnarray}
where $ {\cal M}^{ij}_{\mDRED}(a_1\ldots a_n)={\cal
M}^{ij}_{\mFDH}(a_1\ldots a_n)$ and ${\cal M}_\mNS^{(1)}(a_1\ldots
a_n) $ is the \RS\ independent term appearing also in \Eqn{Mvirtual*}.

We recall that ${\cal M}^{(1)}_\mDRED$ denotes the fully renormalized
one-loop matrix element and \Eqn{Mvirtual} does not contain any
ultraviolet singularities. In \DRED\ this implies that off-shell Green
functions are finite also if external $\gtilde$ are present. This
requires that couplings involving $\ghat$ and couplings involving
$\gtilde$ in general renormalize differently~\cite{Jack:1993ws}. 
As a result, the renormalization procedure in non-supersymmetric
theories can be slightly more involved in \DRED.  We refer to
Appendix~\ref{app:examples} for examples.

\subsection{Real corrections} \label{sec:realDRED}

The calculation of the real corrections in \DRED\ follows the same
pattern as in the other schemes discussed in
Section~\ref{sec:real}. However, there are some important differences
which we will consider for the three singular regions in turn.

\medskip
\emph{Soft Region:} In close analogy to \Eqn{SoftLim}, we have to
consider the limit of the real matrix element when a gluon becomes
soft. In \DRED\ this soft gluon can be either a $\ghat$ or a
$\gtilde$.  The soft limit is governed by eikonal factors of the form
$p_i^\mu p_j^\nu/(s_{ik}s_{jk})$, contracted with the corresponding
polarization sum of the soft gluon $\gth_k$. The polarization sums in
\Eqn{polsumsgphi} then show that a soft $\ghat_k$ leads to the same
limit as a soft full gluon $g_k$, while a soft $\epsilon$-scalar
$\gtilde_k$ leads to zero. Hence,
\begin{align}
{\cal M}^{(0)}_{\mDRED}(a_1,a_2;\ldots g_{k}(p_{k})\ldots 
{\bar a}_{n+1})
&
\stackrel{p_{k}\to 0}{=}
g^2_s\,  \sum_{i,j}
\frac{s_{ij}}{s_{ik} s_{jk}}\, 
{\cal M}^{ij}_{\mDRED}(a_1\ldots a_n) ,
\label{SoftLimDRED}
\\
{\cal M}^{(0)}_{\mDRED}(a_1,a_2;\ldots \gtilde_{k}(p_{k})\ldots 
{\bar a}_{n+1})
&
\stackrel{p_{k}\to 0}{=}
0.
\end{align}
In analogy to the behaviour of ${\cal V}(i,j)$ in the virtual
corrections, the soft limit \Eqn{SoftLimDRED} does not require the
split $g_k=\ghat_k+\gtilde_k$.  Thus, with respect to the soft limit,
\DRED\ is equivalent to \HV\ and \FDH.

\medskip
\emph{Final-State Collinear Region:} Here the split
$g=\ghat + \gtilde$ is essential. According to the main result of
Ref.~\cite{Signer:2005iu}, the key equation for the collinear limit,
\Eqn{CollLim}, has to be modified in \DRED\ if the
parent parton $a_{(kl)}$ is a gluon. In this case the flavour of the
parent parton is not uniquely determined by $\bar a_k$ and $\bar a_l$
and we have to sum over the two possibilities $\ath_{(kl)}
\in\{\ghat,\gtilde\}$.  Thus, in \DRED\ \Eqn{CollLim} becomes
\begin{eqnarray}
\lefteqn{{\cal M}^{(0)}_{\mDRED}
    (a_1,a_2;\ldots \bar a_l(p_l)\ldots \bar
  a_k(p_k)\ldots \bar a_{n+1}) 
\stackrel{p_k\| p_l}{=} } 
\label{CollLimDRED} \\
&& \frac{2\, g_s^2}{s_{kl}} \sum_{\ath_{(kl)}}
P_{{(kl)}^*\to k l}^{<\, \mDRED}(z)\, 
{\cal M}^{(0)}_{\mDRED}(a_1,a_2;\ldots
\ath_{(kl)}(p_k+p_l)\ldots a_{n}) .
\nonumber
\end{eqnarray}
We remark that \Eqn{CollLimDRED} leads us to consider splitting
functions with $\ath_{(kl)}=\ghat$ and $\ath_{(kl)}=\gtilde$ in
\DRED. They are given in Appendix~\ref{app:split}.  \Eqn{CollLimDRED}
remains true if all partons $a_j$ are replaced by $\ath_j$, however,
only for $\ath_i = \ath_{(kl)}$ the split is strictly required. Thus,
\Eqn{outColl} is modified to
\begin{eqnarray}
\label{outCollDRED}
d\sigma^{{\rm real},i}_{\mDRED}(a_1\ldots a_i\ldots a_n)  &=&
  - \frac{\alpha_s}{2\pi} \frac{c_\Gamma}{\eps}\,
   \, \sum_{\ath_i}\, 
   d\sigma^{(0)}_{\mDRED}(a_1\ldots \ath_i\ldots a_n)
\\
\nonumber
&\times&
\sum_{a_k}
\int dz\, \Theta\left(z-\frac{1}{2}\right) 
\frac{(1-z)}{(1-z)_+} P_{i^*\to kl}^{<\, \mDRED}(z)\, .
\end{eqnarray}
This is illustrated in Figure~\ref{fig:outB}, where we consider the
particularly interesting case of an outgoing splitting gluon. For each
choice $\ath_i\in\{\ghat,\gtilde \}$ we have to sum over all splittings
$a_k\in\{g,q,\bar q\}$. Thus, the two terms of Figure~\ref{fig:outA}
become the four terms of Figure~\ref{fig:outB}.

\begin{figure}[t]
\begin{center}
\begin{picture}(400,80)(0,-10)
\SetOffset(20,0)
\Text(-12,45)[l]{$\gamma_{\mDRED}(\ghat)\big|_{N_c}$}
\ZigZag(15,15)(44.6,30){2}{6}
\Gluon(44.6,30)(65.9,51.2){2}{4}
\Gluon(44.6,30)(73.6,37.8){2}{4}
\Line(10,10)(37.2,-2.7)
\Line(10,10)(39.9,7.4)
\Line(10,10)(29.3,-13.)
\GOval(10,10)(10,10)(0){0.8}
\SetOffset(120,0)
\Text(-15,15)[r]{$+$}
\Text(-12,45)[l]{$\gamma_{\mDRED}(\ghat)\big|_{N_F}$}
\ZigZag(15,15)(44.6,30){2}{6}
\ArrowLine(44.6,30)(65.9,51.2)
\ArrowLine(73.6,37.8)(44.6,30)
\Line(10,10)(37.2,-2.7)
\Line(10,10)(39.9,7.4)
\Line(10,10)(29.3,-13.)
\GOval(10,10)(10,10)(0){0.8}
\SetOffset(220,0)
\Text(-15,15)[r]{$+$}
\Text(-12,45)[l]{$\gamma_{\mDRED}(\gtilde)\big|_{N_c}$}
\DashLine(15,15)(44.6,30){2.}
\Gluon(44.6,30)(65.9,51.2){2}{4}
\Gluon(44.6,30)(73.6,37.8){2}{4}
\Line(10,10)(37.2,-2.7)
\Line(10,10)(39.9,7.4)
\Line(10,10)(29.3,-13.)
\GOval(10,10)(10,10)(0){0.8}
\SetOffset(320,0)
\Text(-15,15)[r]{$+$}
\Text(-12,45)[l]{$\gamma_{\mDRED}(\gtilde)\big|_{N_F}$}
\DashLine(15,15)(44.6,30){2.}
\ArrowLine(44.6,30)(65.9,51.2)
\ArrowLine(73.6,37.8)(44.6,30)
\Line(10,10)(37.2,-2.7)
\Line(10,10)(39.9,7.4)
\Line(10,10)(29.3,-13.)
\GOval(10,10)(10,10)(0){0.8}
\end{picture}
\end{center}
\ccaption{}{Illustration of \Eqn{outCollDRED} for the case of  an
outgoing splitting gluon.  The sum over all relevant real processes,
$\sum_{a_k}$, together with the sum due to the split
$\ath_i\in\{\ghat,\gtilde\}$ gives rise to four terms, resulting in
the $N_c$ and $N_F$ parts of $\gamma_{\mDRED}(\ghat)$ and
$\gamma_{\mDRED}(\gtilde)$ respectively. Gluons $g$ and (anti)quarks
are drawn as usual. Dashed lines represent $\gtilde$ and $\ghat$ is
represented by a zigzag line.
\label{fig:outB}}
\end{figure}

\medskip
\emph{Initial-State Collinear Region:} As for the final-state
collinear region, the key point is the necessary split
$g=\ghat+\gtilde$ in the factorization of the collinear limit. In
\DRED, \Eqn{InCollLim} has to be generalized to
\begin{eqnarray}
\lefteqn{
\langle{\cal M}^{(0)}_{\mDRED}(a_1(p_1), a_2;\ldots \bar
  a_{k}(p_k) \ldots \bar a_{n+1})  
\rangle \stackrel{p_k\| p_1}{=} } 
\label{InCollLimDRED} \\
&& \frac{2\, g_s^2}{s_{1k}}  \sum_{\ath_{(1k)}}
P_{1\to {(1k)}^*k}^{<\, \mDRED}(z)\,
\langle{\cal M}^{(0)}_{\mDRED}(\ath_{(1k)}(z\, p_1), a_2;\ldots a_{n})
\rangle .
\nonumber
\end{eqnarray}
Again, as far as the collinear limit is concerned, not all gluons 
in  \Eqn{InCollLimDRED} 
have to be split. Only for the
virtual parton $\ath_{(1k)}$ the split is essential.

Note that in \DRED\ the crossed splitting functions satisfy the
crossing relation \Eqn{CrossingViolation} without $\Delta$ terms.
Hence, the initial state collinear term for parton 1 can be written as
\begin{eqnarray}
\label{inCollDRED}
d\sigma^{{\rm real},1}_{\mDRED}(a_1\ldots a_n) &=& 
   \frac{\alpha_s}{2\pi} \frac{c_\Gamma}{\eps}
\bigg[\sum_{\ath_1}\gamma_{\mDRED}(\ath_1)\, 
d\sigma^{(0)}_{\mDRED}(\ath_1(p_1),a_2;\ldots a_n)  
\\ \nonumber
&& \hspace*{-2cm}-\  \sum_{a_k} \sum_{\ath_{(1k)}} \int dz\
 P_{1\to (1k) k}^{\mDRED}\
d\sigma^{(0)}_{\mDRED}(\ath_{(1k)}(z\, p_1),a_2;\ldots a_n) \bigg] .
\end{eqnarray}
As illustrated in Figure~\ref{fig:inB}, in the sum over $\ath_{(1k)}$
in \Eqn{inCollDRED} it is essential that we treat $\ghat$ and
$\gtilde$ as separate partons, whereas $a_k\in\{g,q,\bar q\}$.
For an incoming gluon, the three terms of Figure~\ref{fig:inA} are
generalized in \DRED\ to the four terms of  Figure~\ref{fig:inB}.
 
\begin{figure}[t]
\begin{center}
\begin{picture}(400,60)(0,-10)
\Text(85,15)[l]{$+$}
\Text(190,15)[l]{$+$}
\Text(295,15)[l]{$+$}
\SetOffset(-15,0)
\ZigZag(65,15)(36.4,20.4){2}{6}
\Gluon(36.4,20.4)(15.0,20.4){2}{3}
\Gluon(36.4,20.4)(89.3,36.8){2}{8}
\Line(70,10)(97.2,-2.7)
\Line(70,10)(91.2,-11.2)
\Line(70,7)(31.4,-3.4)
\GOval(70,10)(10,10)(0){0.8}
\SetOffset(90,0)
\DashLine(65,15)(36.4,20.4){2.}
\Gluon(36.4,20.4)(15.0,20.4){2}{3}
\Gluon(36.4,20.4)(89.3,36.8){2}{8}
\Line(70,10)(97.2,-2.7)
\Line(70,10)(91.2,-11.2)
\Line(70,7)(31.4,-3.4)
\GOval(70,10)(10,10)(0){0.8}
\SetOffset(195,0)
\Gluon(36.4,20.4)(15.0,20.4){2}{3}
\ArrowLine(65,15)(36.4,20.4)
\ArrowLine(36.4,20.4)(89.3,36.8)
\Line(70,10)(97.2,-2.7)
\Line(70,10)(91.2,-11.2)
\Line(70,7)(31.4,-3.4)
\GOval(70,10)(10,10)(0){0.8}
\SetOffset(300,0)
\Gluon(36.4,20.4)(15.0,20.4){2}{3}
\ArrowLine(36.4,20.4)(65,15)
\ArrowLine(89.3,36.8)(36.4,20.4)
\Line(70,10)(97.2,-2.7)
\Line(70,10)(91.2,-11.2)
\Line(70,7)(31.4,-3.4)
\GOval(70,10)(10,10)(0){0.8}
\end{picture}
\end{center}
\ccaption{}{Illustration of \Eqn{inCollDRED} for an incoming splitting
  gluon with partons represented as in Figure~\ref{fig:outB}. The sum
  over all relevant real processes, $\sum_{a_k}$, together with the
  sum due to the split $\ath_{(1k)}\in\{\ghat,\gtilde\}$ gives rise to
  four terms.  \label{fig:inB}}
\end{figure}

\subsection{Collinear counterterm} \label{sec:incollDRED}

The collinear counterterm in \DRED\ can now be constructed in the same
way as in the other schemes. Generalizing \Eqn{NonMScoll} to \DRED\ we
can write
\begin{eqnarray}
\label{NonMScollDRED}
\lefteqn{d\sigma^{\rm coll}_{\mDRED,\mFS}(a_1,a_2;\ldots a_n) =
\frac{\alpha_s}{2\pi} \frac{c_\Gamma}{\eps} \sum_{a_k} \sum_{\ath_i} 
\int dz} && 
\\ \nonumber
&\times& \bigg[ 
\left(P_{1\to i k}^{\mDRED}(z)
 + \epsilon X^{\mFS}_{1\to ik}(z) \right) \
d\sigma^{(0)}_{\mDRED}(\ath_i(z\, p_1),a_2(p_2);\ldots a_n) 
 \\ \nonumber
&& + \,  
\left(P_{2\to i k}^{\mDRED}(z)
 + \epsilon X^{\mFS}_{2\to ik}(z) \right)\
d\sigma^{(0)}_{\mDRED}(a_1(p_1),\ath_i(z\, p_2);\ldots a_n) \bigg]\ .
\end{eqnarray}
The $X^{\mFS}_{l\to ik}$ involving partons
$\ath_i\in\{\ghat,\gtilde\}$ appearing here are defined in terms of
the functions appearing in \Eqn{NonMScoll} as 
\begin{equation}
\label{XDREDdef}
X^{\mFS}_{l\to\gth k}=
\frac{ \omega_{\mDRED}(\gth)}{\omega_{\mDRED}(g)} \,
X^{\mFS}_{l\to g k} \ .
\end{equation}
With these definitions, the results of this section,
\Eqns{MvirtualFull}{inCollDRED}, show that the hard
partonic cross section
\begin{eqnarray}
\label{subNonMSDRED}
\lefteqn{d\hat\sigma_{\mFS}(a_1\ldots a_n) =
\Big[ d\sigma^{\rm Born}_{\mDRED} (a_1\ldots a_n) } &&
\\ \nonumber &&
+\ d\sigma^{\rm virt}_{\mDRED}(a_1\ldots a_n)
+ d\sigma^{\rm real}_{\mDRED}(a_1\ldots a_n)
+ d\sigma^{\rm coll}_{\mDRED,\mFS}(a_1\ldots a_n) \Big]_{D\to 4}
\end{eqnarray}
is equal to the one in the other schemes given in \Eqn{subNonMS}. 
This shows in particular that it is possible to realize the
\MSbar\ factorization scheme in \DRED\ in the same way as in \CDR,
\HV, or the \FDH\ scheme. In order to make this result explicit we
close the section by providing the full form of the appropriate
collinear counterterm, valid in all \RS,
\begin{eqnarray}
\label{collctX}
\lefteqn{d\sigma^{\rm coll}_{\mRS,\mMS}(a_1,a_2;\ldots a_n) = 
\frac{\alpha_s}{2\pi} \frac{c_\Gamma}{\eps} 
\sum_{a_k} \sum_{\ath_i} \int dz}
&&
\\ \nonumber &&
\times
\bigg[
\left( P_{1\to i k}^{\mRS}(z) - 
 P_{1\to i k}^{\mMS\,\eps}(z) \right)
d\sigma^{(0)}_\mRS(\ath_i(z\, p_1),a_2;\ldots )
+\{1\leftrightarrow 2\}
\bigg]
,
\end{eqnarray}
where the sum $\sum_{a_k}$ runs over $a_k\in\{g,q,\bar{q}\}$ in all
\RS, whereas the sum $\sum_{\ath_i}$ runs over in
$\ath_i\in\{g,q,\bar{q}\}$ in \CDR,\HV, \FDH, and over
$\ath_i\in\{\ghat,\gtilde,q,\bar{q}\}$ in \DRED. Also, in \DRED\ we
define\footnote{%
The $\Delta$ terms appearing in the \HV\ and  \FDH\ schemes, see
\Eqns{inColl}{NonMScoll}, have been ignored. According to the remark
at the end of Section~\ref{sec:incoll} this is correct if the $\Delta$
terms are also ignored in the real corrections.}
\begin{equation}
\label{PDREDdef}
P_{1\to \gth k}^{\mMS\, \eps}(z) \equiv
\frac{\omega_{\mDRED}(\gth)}{\omega_{\mDRED}(g)} \,
P_{1\to g k}^{\mMS\, \eps}(z) 
\end{equation}
in analogy to \Eqn{XDREDdef}.

\section{Applying {DRED}} \label{sec:spxs}

In the previous two sections we discussed how the singularities and
\RS\ dependence between the various parts of \Eqn{sigmahat} cancel, and
we found that the subtracted partonic cross sections given in
\Eqns{subNonMS}{subNonMSDRED} are finite and \RS\
independent. This is precisely what we wanted to achieve. However,
there is still one conceptual issue to be addressed. 

The question is whether in the convolution of subtracted partonic
cross sections with PDF we need to distinguish between $\ghat$ and
$\gtilde$ in DRED. We will show that this is not the case. 
 This will also entail that no PDF for finding an unphysical
$\gtilde$ in a hadron will be required.

Once this issue is clarified, we will summarize our results
and give transition rules between the various \RS\ separately for all
parts of the subtracted finite partonic cross sections.

\subsection{Parton distribution functions in DRED} 
\label{sec:combine}

The results \Eqns{subNonMS}{subNonMSDRED} have been given without
taking into account that the subtracted partonic cross sections have
to be multiplied by PDF. In \DRED\ it might seem natural to
distinguish partonic cross sections with $\ghat$ or  $\gtilde$ in the
initial state and convolute them with different PDF. We will show
that this is not required and that we can use \Eqn{subNonMSDRED} for
initial state full gluons $g$, convoluted with just one PDF even in
DRED.  In particular, there is no need to introduce unphysical PDF
for finding a $\gtilde$ in a hadron.

In the strict spirit of \DRED\ it is correct to consider independent
PDF for $\ghat$ and $\gtilde$ and write a hadronic cross section as a
sum of the form
\begin{equation}
f_{\ghat/H} \otimes
d\hat\sigma_{\mFS}(\ghat_1\ldots)
+
f_{\gtilde/H} \otimes
d\hat\sigma_{\mFS}(\gtilde_1\ldots) 
+
f_{q/H} \otimes
d\hat\sigma_{\mFS}(q_1\ldots) 
+
f_{\bar q/H} \otimes
d\hat\sigma_{\mFS}(\bar q_1\ldots) .
\label{HadDRED}
\end{equation}
The partonic cross sections $d\hat\sigma(\gth_1)$ can be constructed
in the same way as 
\Eqn{subNonMSDRED}. They are individually finite and satisfy
\begin{equation}
\sum_{\ath_1} \sum_{\ath_2}
\frac{\omega_\mDRED(\ath_1)}{\omega_\mDRED(a_1)}
\frac{\omega_\mDRED(\ath_2)}{\omega_\mDRED(a_2)}\
d\hat\sigma_{\mFS}(\ath_1,\ath_2;\ldots a_n) = 
d\hat\sigma_{\mFS}(a_1,a_2;\ldots a_n) .
\label{subNonMSDREDsum}
\end{equation}
All the PDF in \Eqn{HadDRED} would be obtained by performing a fit at
one particular factorization scale $\mu_0$ and then using
Altarelli-Parisi equations to evolve them to any other scale $\mu$.

The central point is that the unphysical PDF $f_{\gtilde/H}$ is of the
order $\epsilon$, and hence its contributions to both the hadronic
cross section, \Eqn{HadDRED}, and to the evolution of the other PDF
are of the order $\epsilon$ and thus negligible.

In order to prove this we start by noting that since in other
regularization schemes one gluon PDF $f_{g/H}$ is sufficient it is
possible to arrange the fit in \DRED\ such that
$f_{\gtilde/H}(\mu_0)=0$ at the starting scale $\mu_0$. The evolution
is given by the Altarelli-Parisi equations, generalized to include
$\gtilde$:
\begin{equation}
\mu^2 \frac{\partial}{\partial \mu^2} 
\left( \begin{array}{c} 
      f_{q/H}(z) \\ f_{\ghat/H}(z) \\ f_{\gtilde/H}(z) 
     \end{array} \right) 
= \frac{\alpha_s}{2\pi} \int_{z}^1 \frac{d\xi}{\xi}
\left( \begin{array}{ccc} 
   P_{q\to q} & P_{\ghat\to q} & P_{\gtilde \to q}\\
   P_{q\to \ghat} & P_{\ghat\to \ghat} & P_{\gtilde \to \ghat}\\
   P_{q\to \gtilde} & P_{\ghat\to \gtilde} & P_{\gtilde \to \gtilde}\\
\end{array} \right)
\left( \begin{array}{c} 
      f_{q/H}(\xi) \\ f_{\ghat/H}(\xi) \\ f_{\gtilde/H}(\xi) 
     \end{array} \right) ,
\end{equation}
where we have suppressed the $\mu$ dependence of $\alpha_s$ and
$f_{a_i/H}$ and  have used the short-hand notation $P_{i\to j} \equiv
P^\mDRED_{i\to j (ij)}(z/\xi)$.  The evolution of 
$f_{\gtilde/H}(z)$ gets contributions from $P_{q\to
\gtilde}\times f_{q/H}$, $P_{\ghat\to \gtilde} \times f_{\ghat/H}$ and
$P_{\gtilde \to \gtilde} \times f_{\gtilde/H}$. They are ${\cal
O}(\eps) \times 1$, ${\cal O}(\eps) \times 1$ and $1\times {\cal
O}(\eps)$, respectively, confirming that $f_{\gtilde/H}(\mu) = {\cal
O}(\eps)$ for all scales $\mu$. This in turn implies that the
contribution to the evolution of $f_{q/H}$ and $f_{\ghat/H}$ due to
$f_{\gtilde/H}$ is also ${\cal O}(\eps)$. The situation is in fact
very similar to the contributions due to quarks if there were
$N_F=\eps$ flavours. Finally, $d\hat\sigma_{\mFS}(\ath_1,\ath_2;\ldots
a_n)$ as appearing in \Eqns{HadDRED}{subNonMSDREDsum} is finite for
all initial states $\ath_1,\ath_2$ separately.  Hence the contribution
of the unphysical PDF $f_{\gtilde/H}$ in \Eqn{HadDRED} is ${\cal
O}(\epsilon)$.

This confirms that \DRED\ can be used throughout all parts of
calculations for hadronic cross sections at one loop, without the need
for unphysical PDF. It is correct to consider only one gluon PDF
$f_{g/H}$ and the combined cross sections $d\hat\sigma(a_1,a_2;\ldots
a_n)$ from \Eqn{subNonMSDRED} also in \DRED. In other words,
\Eqn{SigmaHad} is correct in all schemes, including \DRED, if the sums
over all parton types include only the full gluon $g$ and quarks $q$,
$\bar{q}$ (and possibly further, massive partons).

\subsection{Summary of practical computations in DRED} \label{sec:summary}

Let us finally summarize how to do a next-to-leading order calculation
in practice in \DRED\ or any other scheme. The main point is that only
for the ultraviolet renormalization and the collinear counterterm a
split of $g$ into $\ghat+\gtilde$ is required.

\medskip
\emph{Virtual Corrections:} To obtain the virtual corrections we start
by computing ${\cal M}^{(1)}_\mRS(a_1\ldots a_n)$ with $a_i
\in\{g,q,\bar q\}$. For the actual calculation of the one-loop diagrams we
do not need to split the process into many different parts as in
\Eqn{Mvirtual}, but can compute directly with $g$. 
The structure of
the ultraviolet counterterms depends on the \RS\ and the symmetries of
the underlying theory. If \DRED\ is used in a supersymmetric context,
even the counterterms can typically be computed without the split, and
usual multiplicative renormalization is sufficient to generate the
counterterms. Using \HV\ or \CDR\ in supersymmetric theories leads to
the complication of non-multiplicative, supersymmetry-restoring
counterterms. In non-supersymmetric theories, determining counterterms
in \DRED\ requires the split 
$g=\ghat+\gtilde$. 

Once we have the renormalized
one-loop matrix element, ${\cal M}^{(1)}_\mRS(a_1\ldots a_n)$ we
obtain $d\sigma^{\rm virt}_\mRS$ by integration over the phase
space. Again, a split of $d\sigma^{\rm virt}_\mRS$ as in
\Eqn{subNonMSDREDsum} is not required.  The split $g=\ghat+\gtilde$
for external gluons becomes useful if we want to express the
singularity structure or \RS\ dependence in a simple way as done in
the next subsection.
\medskip

\emph{Real Corrections:} The real corrections in \DRED\ can be
obtained in a straightforward way by directly integrating the
4-dimensional tree-level matrix elements containing only 4-dimensional
gluons $g$ and (anti)quarks. Likewise, in \CDR\ we have to integrate
the $D$-dimensional tree-level matrix elements. Regarding the real
corrections in \HV\ and \FDH\ we remind the reader of a subtlety
related to unitarity (see also Ref.\ \cite{Catani:1996pk}). At first
sight it might appear that there is no 
difference in $d\sigma_{\mRS}^{\rm real}$ for schemes where the
tree-level matrix elements are evaluated in four dimensions,
e.g. between $d\sigma_{\mHV}^{\rm real}$ and $d\sigma_{\mFDH}^{\rm
real}$. However, this is not correct. In order to maintain unitarity,
in the singular regions initial and final state partons have to be
treated in the same way as 
 partons in a closed loop, i.e. as ``internal''. Thus
in \HV\ and \FDH\ it is not correct to simply integrate the
corresponding four-dimensional real tree-level matrix elements over
the phase space.  In particular, \Eqns{CollLim}{InCollLim} contain
${\cal O}(\eps)$ terms in \FDH\ and \HV\ and result in finite
differences between $d\sigma_{\mHV}^{\rm real}$ and
$d\sigma_{\mFDH}^{\rm real}$, even though the tree-level matrix
elements agree.

In principle this procedure leads to the $\Delta$ terms in
\Eqn{inColl} because in the \HV\ and \FDH\ schemes incoming and
outgoing splittings differ. As discussed in Section~\ref{sec:incoll} it
is possible to redefine the results by ignoring the $\Delta$ terms if
the same is done in the collinear counterterms.

\medskip

\emph{Collinear Counterterm:} The collinear counterterm 
is given by \Eqn{collctX}. It realizes the \MSbar\ factorization
scheme independent of the \RS\ used for the computation.  For \DRED,
we stress that for the term given explicitly on the r.h.s. of
\Eqn{collctX} the  partons $a_1\ldots a_n$ as well as $a_k$
are never $\ghat$ or $\gtilde$ separately, but can always be combined
to $g$. For the virtual parton $\ath_i$, however, it is important to
treat $\ghat$ and $\gtilde$ separately~\cite{Signer:2005iu}. This
requires the use of the splitting functions given in
\Eqns{PdredH}{PdredT}.  For the \HV\ and \FDH\ schemes, the $\Delta$
terms can be ignored in accordance with the computation of the real
corrections.

\subsection{Translation rules between  different schemes}
\label{sec:SchemeRelations}

In the following we describe how the results in the various \RS\ are
related, making use of the split $g=\ghat+\gtilde$ as appropriate. We
will focus on the virtual corrections; similar results for the real
corrections and the collinear counterterms can be trivially obtained
from Eqns.\ (\ref{outCollDRED}), (\ref{inCollDRED}),
(\ref{collctX}). Starting from our  
renormalized result in \CDR, ${\cal M}^{(1)}_\mCDR(a_1\ldots a_n)$,
written in the form of
\Eqn{Mvirtual*}, we can obtain ${\cal M}^{(1)}_\mHV(a_1\ldots a_n)$
simply by replacing the $D$ dimensional (colour-linked) Born terms by
the corresponding 4-dimensional expressions. No further change is
required, since $\gamma_\mCDR(a_i) = \gamma_\mHV(a_i)$.

The only
difference between the \HV\ and \FDH\ scheme on the other hand does
come from the differences of the $\gamma_\mRS$ terms, which have been
explained in Section~\ref{sec:splittings} and quantitatively given in
\Eqn{gammaALL}. For a process with $\#g$ gluons, $\#q$ massless
(anti)quarks and $\#Q$ massive (anti)quarks the difference is
\begin{eqnarray}
\lefteqn{{\cal M}^{(1)}_\mFDH(a_1\ldots a_n) - 
  {\cal M}^{(1)}_\mHV(a_1\ldots a_n)} &&
 \label{MVfhd-hv}\\
&=&\frac{\alpha_s}{2\pi}\, {\cal M}^{(0)}_\mFDH(a_1\ldots a_n)\, 
\sum_{x\in\{g,q,Q\}} \, \#x 
\left(\frac{\gamma_\mHV(a_x)-\gamma_\mFDH(a_x)}{\eps} \right) \nonumber \\
&=&
\frac{\alpha_s}{2\pi}\, {\cal M}^{(0)}_\mFDH(a_1\ldots a_n)\, 
\left[ \#g\, \frac{N_c}{6} + \#q\, \frac{C_F}{2} \right] .
\nonumber
\end{eqnarray}
In the second line the influence of the different $\gamma_\mRS$ for
all parton types is made explicit, in the third line the result in
brought into a compact form.

The difference between \FDH\ and \DRED\ is obtained by taking the
difference between \Eqns{MvirtualFull}{Mvirtual*}.
We can bring it
into a
simple form by using that we can write the tree-level quantities in
\FDH\ in a \DRED-like form,
\begin{equation}
{\cal M}^{(0)}_\mFDH(a_1\ldots a_i \ldots a_n) = 
\sum_{\ath_i}{\cal M}^{(0)}_\mDRED(a_1\ldots \ath_i\ldots a_n) .
\end{equation}
The difference is then governed by the factors 
$\gamma_\mDRED(\ghat)- \gamma_\mFDH(g)$ and $\gamma_\mDRED(\gtilde)
- \gamma_\mFDH(g)$. As explained in Section \ref{sec:splittings} the
first of these vanishes. The second is non-zero,
\begin{equation}
\label{gammacombine}
\gamma_\mDRED(\gtilde)
- \gamma_\mFDH(g)= \frac{1+\epsilon}{6} \left(N_c-2\, T_F N_F\right)
,
\end{equation}
and is present for every $\gtilde$ in the initial or final state, see
\Eqn{MvirtualFull}.
Exploiting also that at leading order $\alpha_s=\alpha_e$, we obtain
\begin{eqnarray}
\lefteqn{{\cal M}^{(1)}_\mDRED(a_1\ldots a_n) - 
         {\cal M}^{(1)}_\mFDH(a_1\ldots a_n)} &&
\label{MVfhd-dred} \\
&=& 
\frac{\alpha_s}{2\pi}\, 
\frac{\gamma_\mFDH(g)-\gamma_\mDRED(\gtilde)}{\eps} 
\sum_{\{\ath\}}\, \#\gtilde(\{\ath\})\, 
{\cal M}^{(0)}_\mDRED(\ath_1\ldots \ath_n) 
\nonumber \\
&=& 
\frac{\alpha_s}{2\pi}\, \frac{1}{\eps} \, \frac{2\, T_F N_F-N_c}{6}
\sum_{j = 1}^{\#g}\,  
{\cal M}^{(0)}_\mDRED(a_1\ldots a_n) 
\big|_{g_{i_j} \to \gtilde_{i_j}} .
\nonumber
\end{eqnarray}
Again, in the second line the influence of the different $\gamma_\mRS$
relevant for $\gtilde$ is made explicit. In the third line we used that for
processes with at least one $\gtilde$, i.e. with $\#\gtilde(\{\ath\})\ge 1$ 
we have $ {\cal M}^{(0)}_\mDRED
\sim \eps$; therefore we neglected the   ${\cal O}(\eps)$ terms from
\Eqn{gammacombine}. The notation ${\cal   M}^{(0)}_\mDRED(a_1\ldots 
a_n)\big|_{g_{i_j} \to \gtilde_{i_j}}$ implies that all gluons except
gluon $i_j$ are 4-dimensional gluons. Thus in the final expression on
the r.h.s. of \Eqn{MVfhd-dred} we sum over all processes where one
4-dimensional gluon $g$ at a time is replaced by a $\gtilde$.

The transition rules \Eqns{MVfhd-hv}{MVfhd-dred} ignore ${\cal
O}(\eps)$ terms and are given for pure QCD processes. However, they
can easily be generalized to other processes, involving e.g.\ photons
or massive partons,  simply by using the
corresponding explicit expressions for $\gamma_\mRS$.

\section{Conclusions} \label{sec:conclusions}

The main result presented in this paper is that \DRED\ can be used for
the calculation of cross sections at NLO, even for processes with
hadrons in the initial state. Problems related to factorization, as
reported in the literature~\cite{Beenakker:1988bq, Beenakker:1996dw,
Smith:2004ck}, can be avoided by taking into account the appropriate,
generalized factorization in \DRED~\cite{Signer:2005iu}. We have shown
explicitly how to use \DRED\ together with an arbitrary factorization
scheme. In particular, the conventional PDF~\cite{MRST,CTEQ} in the
\MSbar-factorization scheme can be used. Also we have given explicit rules
on how to transform separately the various parts of the hard partonic
cross section, \Eqn{sigmahat}, from \DRED\ to other \RS. This
completes the previously known set of transition rules between \CDR,
\HV\ and \FDH~\cite{Kunszt:1993sd, Catani:2000ef, Catani:1996pk}. It
is thus possible to use different \RS\ for different parts of the
calculation which might help simplifying the explicit computations. In
this context we also reiterate the distinction between \FDH\ and
\DRED. According to the definitions of the \RS\ given in
Section~\ref{sec:DR}, at one loop \FDH\ is equivalent to the scheme
\DR\ used e.g. in Refs.~\cite{Kunszt:1993sd, Catani:2000ef,
Catani:1996pk} but differs from \DRED\ used e.g. in
Refs.~\cite{Siegel79,CJN80,DS05}.

The salient feature of a consistent use of \DRED\ is the split
$g=\ghat+\gtilde$. In practice, this split does not significantly
complicate calculations. It is needed mainly for the correct treatment
of the collinear limit of squared matrix elements. Thus it affects the
collinear counterterm and the phase-space integration over the
singular, collinear region. The modifications regarding the former are
shown in \Eqns{inColl}{inCollDRED}. For the phase-space integration,
the usual procedures have to be slightly modified. For the method
presented in Ref.~\cite{Frixione:1995ms} for example, the collinear
singularities in the real corrections are singled out using
distributions, enforcing the collinear limit of the real matrix
element squared. If this method is to be used together with \DRED\
this simply means that the proper collinear limit,
\Eqns{CollLimDRED}{InCollLimDRED}, has to be taken. For the dipole
subtraction method~\cite{Catani:1996vz} additional dipoles with
$\gtilde$ are required. These can be obtained making minor
modifications of existing dipoles, similar to the corresponding
adaptation to \FDH~\cite{Campbell:2004ch}.

In some cases, the split $g=\ghat+\gtilde$ is also  required for the
ultraviolet 
counterterms, since e.g. the couplings $\ghat q \bar{q}$ and $\gtilde q
\bar{q}$ renormalize differently. This seems to be a disadvantage of
\DRED. On the other hand, one of the advantages of \DRED\ is that in
supersymmetric theories  no
supersymmetry-restoring counterterms are 
required (in many practical cases; for a recent discussion see
Ref.\ \cite{DS05}). In this case, also couplings with $\ghat$ and
$\gtilde$ 
renormalize identically and renormalization is actually simpler in \DRED\
than e.g. in \CDR. This facilitates the use of the \DRbar-scheme for
supersymmetric parameters which is used in a wide variety of
calculations~\cite{AguilarSaavedra:2005pw}. 

Thus, \DRED\ is a
\RS\ which is well compatible with supersymmetry and which can be
realized with minimal 
modifications compared to \CDR\ and used for an arbitrary cross
section at NLO.
In the past, following the examples of
e.g.\ Refs.\ \cite{Beenakker:1996dw,Beenakker:1996ch}, many 
predictions for supersymmetric processes at hadron colliders were
calculated using \CDR\ in spite of the required
supersymmetry-restoring counterterms. In the future, similar
calculations can alternatively be carried out using \DRED, which can
lead to simplifications with the present, better understanding of
\DRED.

\vspace*{0.5em}
\noindent
\subsubsection*{Acknowledgement}
This work is supported in part by the European Community's Marie-Curie
Research Training Network under contract MRTN-CT-2006-035505 `Tools
and Precision Calculations for Physics Discoveries at Colliders'.

\appendix

\section{Collinear limits and sum rules} \label{app:split}

In this appendix we study the collinear limit of squared matrix
elements and derive the associated splitting functions and
$\gamma_\mRS$ terms. Even though most of the results presented here
are well known, we repeat them for the reader's convenience and to
fix our notation and conventions.

Following Ref.~\cite{Catani:1996pk}, we consider a slightly off-shell
massless outgoing parton $a_i(p_i)$ that splits into massless on-shell
partons $a_k(p_k)$ and $a_l(p_l)$. The momenta are parametrized as
\begin{eqnarray}
p_k^\mu &=& z\, p^\mu + k_\perp^\mu 
  - \frac{k_\perp^2}{z}\, \frac{n^\mu}{2\, (p\cdot n)},
\label{pk} \\
p_l^\mu &=& (1-z)\, p^\mu - k_\perp^\mu 
  - \frac{k_\perp^2}{(1-z)}\, \frac{n^\mu}{2\, (p\cdot n)},
\label{pl}
\end{eqnarray}
with $p^2 = n^2 = (k_\perp\cdot p)= (k_\perp\cdot n) = 0$. The
invariant mass of the incoming parton is $p_i^2 = 2 (p_k\cdot p_l) = -
k_\perp^2/(z(1-z))$ and vanishes in the collinear limit $k_\perp^\mu
\to 0$.

To start with we consider the particularly interesting case where the
parent parton is a gluon. Denoting by ${\cal A}_\mu^m$ the amplitude
of the parent process, stripped of its polarization vector
$\varepsilon^\mu(p_i)$, we can write the collinear limit of the full
process as
\begin{equation}
{\cal M}^{(0)}_\mRS(1\ldots g(p_k),g(p_l) \ldots n+1) 
\stackrel{p_k\| p_l}{=} 
\frac{4\pi\, \alpha_s}{p_k\cdot p_l}\,
\delta_{mn}\, \, {\cal A}_\mu^m\,
{\cal P}^{<\, \mRS\, \mu\nu}_{g_i^*\to g_k g_l}\,  
{\cal A}_\nu^{*\, n} \  ,
\label{ggcollA}
\end{equation}
where $m$ and $n$ are colour labels. After averaging over
$k_\perp^\mu$, the operator ${\cal P}^{<\, \mRS\, \mu\nu}_{g_i^*\to
g_k g_l}$ is proportional to the metric, which in accordance with
\Eqn{polsumsgphi} corresponds to the polarization sum $\sum
\varepsilon^\mu \varepsilon^{*\, \nu}$. Explicitly we find
\begin{eqnarray}
{\cal P}^{<\, \mCDR\, \mu\nu}_{g^*\to g g}(z)&=& -\ghat^{\mu\nu}\, (2 N_c)\,
              \left(\frac{z}{1-z} + \frac{1-z}{z} + z (1-z)\right)\ ,
\label{PGtoGGcdr} \\
{\cal P}^{<\, \mHV\, \mu\nu}_{g^*\to g g}(z)&=& 
              - \bar{g}^{\mu\nu}\, (2 N_c)\, 
              \left(\frac{z}{1-z} + \frac{1-z}{z} + z (1-z)\right)\ ,
\label{PGtoGGhv} \\
{\cal P}^{<\, \mFDH\, \mu\nu}_{g^*\to g g}(z)&=&  
              - \bar{g}^{\mu\nu}\, (2 N_c)\,
              \left(\frac{z}{1-z} + \frac{1-z}{z} +
              \frac{2}{D-2}\, z (1-z)\right)\ ,
\label{PGtoGGfdh} \\
{\cal P}^{<\, \mDRED\, \mu\nu}_{g^*\to g g}(z)&=& 
-\ghat^{\mu\nu}\, (2 N_c) 
   \left(\frac{z}{1-z} + \frac{1-z}{z} + \frac{2}{D-2}\, z (1-z)
         \right) 
\label{PGtoGGdred} \\
&&  -\gtilde^{\mu\nu}\, (2 N_c) 
  \left(\frac{z}{1-z} + \frac{1-z}{z}\right)\ .
\nonumber
\end{eqnarray}
The interesting point is that in \DRED\ we get a combination of
$\ghat^{\mu\nu}$ and $\gtilde^{\mu\nu}$. Thus the collinear limit has
to be written as a sum over two terms as in \Eqn{CollLim}. In the
spirit of \DRED, we can further disentangle the splitting
operator and write
\begin{eqnarray}
{\cal P}^{<\,\mDRED\, \mu\nu}_{g^*\to g g} &=&  
{\cal P}^{<\,\mDRED\, \mu\nu}_{\ghat^*\to \ghat \ghat} +
{\cal P}^{<\,\mDRED\, \mu\nu}_{\ghat^* \to \gtilde \gtilde}+
{\cal P}^{<\,\mDRED\, \mu\nu}_{\gtilde^* \to \ghat \gtilde}+
{\cal P}^{<\,\mDRED\, \mu\nu}_{\gtilde^* \to \gtilde \ghat} 
\label{Pdisentangle} \\
&=&-\ghat^{\mu\nu}\, (2 N_c)
              \left(\frac{z}{1-z} + \frac{1-z}{z} + z (1-z)\right)
\nonumber \\
&  &  - \ghat^{\mu\nu}\, (2 N_c)\, \frac{4-D}{D-2}\, z (1-z)
      - \gtilde^{\mu\nu}\, (2 N_c)\, \frac{1-z}{z} 
      - \gtilde^{\mu\nu}\, (2 N_c)\, \frac{z}{1-z} \ .\nonumber
\end{eqnarray}
The splitting functions $P^{<\, \mRS}_{g^*\to g g}$ can be read off of
Eqs.~(\ref{PGtoGGcdr}) -- (\ref{Pdisentangle}) simply by dropping the
polarization sum. Performing similar calculations for all other
possible splits we find the following results:
\begin{align}
P^{<\, \mCDR}_{g^*\to g g}&
 = P^{<\, \mDRED}_{\ghat^*\to \ghat \ghat} 
= 2 N_c\, \left(\frac{z}{1-z} + \frac{1-z}{z} + z (1-z)\right)\ ,
\label{GtoGGcdr} \\
P^{<\, \mFDH}_{g^*\to g g} &= P^{<\, \mDRED}_{\ghat^*\to g g} 
= 2 N_c\,\left(\frac{z}{1-z} + \frac{1-z}{z} + \frac{2}{D-2}\, z (1-z)
         \right)\ ,
\label{GtoGGfdh}  \\
 P^{<\, \mCDR}_{g^*\to q\bar{q}} 
= P^{<\, \mFDH}_{g^*\to q\bar{q}}& 
= P^{<\, \mDRED}_{\ghat^*\to q\bar{q}} 
=T_F\, \left(1-\frac{4}{D-2} z (1-z) \right)\ ,
\label{GtoQQ} \\
 P^{<\, \mCDR}_{q^*\to q g}&
= P^{<\, \mDRED}_{q^*\to q \ghat}
=C_F\, \left(\frac{2z}{1-z}+\frac{D-2}{2} (1-z) \right)\ ,
\label{QtoQGcdr} \\
 P^{<\, \mFDH}_{q^*\to q g} &= P^{<\, \mDRED}_{q^*\to q g} 
=C_F\, \left(\frac{2z}{1-z}+ (1-z) \right) \ .
\label{QtoQGfdh}
\end{align}
The results for \HV\ are always identical to the ones for \CDR, 
\begin{eqnarray}
&& P^{<\, \mCDR}_{i^*\to k l} = P^{<\, \mHV}_{i^*\to k l}. 
\label{PHVCDR}
\end{eqnarray}
The results particular to \DRED\ are given by
\begin{eqnarray}
&& P^{<\, \mDRED}_{\ghat^*\to \gtilde \gtilde}
= 2 N_c\, \frac{4-D}{D-2}\ z (1-z) ,
\label{GhattoGG}\\
&& P^{<\, \mDRED}_{\gtilde^*\to \ghat \gtilde}
= 2 N_c\,\left(\frac{1-z}{z} \right),
\label{GtildetoGG}\\
&& P^{<\, \mDRED}_{\gtilde^*\to q\bar{q}} =  T_F,
\label{GtildetoQQ}\\
&&  P^{<\, \mDRED}_{q^*\to q \gtilde}
=C_F\, \frac{4-D}{2}\ (1-z)  .
\label{QtoQGdred} 
\end{eqnarray}
The remaining splitting functions can be obtained by $P^{<\,
\mRS}_{i^*\to kl}(z) = P^{<\, \mRS}_{i^*\to lk}(1-z)$. The splitting
functions appropriate for the split of an incoming parton $P^{<\,
\mRS}_{i\to k^*l}(z)$ can be obtained through the crossing relation
\Eqn{Pcrossing} with
\begin{align}
\nonumber
\omega_\mHV(g) = \omega_\mFDH(g) = \omega_\mDRED(g) &= 2\, (N_c^2-1),
&&\omega_\mRS(q) = 2\, N_c ,
\\
\nonumber
\omega_\mCDR(g) = \omega_\mDRED(\ghat) &= (D-2) (N_c^2-1)
,
\\
\omega_\mDRED(\gtilde) &= (4-D)(N_c^2-1)
\label{omegas}
. 
\end{align}
We note that in \CDR\ and in \DRED\ there is no difference between
the splitting functions for incoming and outgoing partons,
i.e.\ \Eqn{CrossingViolation} holds with all $\Delta$ terms equal to
zero. Thus, \Eqns{GhattoGG}{GtildetoGG} and
\Eqns{GtildetoQQ}{QtoQGdred} are not independent. For the \HV\ and
\FDH\ scheme the $\Delta$ terms do not vanish. In the \FDH\ scheme,
for example, the different coefficient of the three terms on the
r.h.s.\ of \Eqn{GtoGGfdh} lead to $\Delta^{\mFDH}_{g\to gg}\ne0$. In
the \HV\ scheme, the different $\omega_\mHV(g)\ne\omega_\mCDR(g)$,
together with \Eqn{PHVCDR}, are the origin of the non-vanishing
$\Delta^\mHV$ terms.

The splitting functions $P^{<\, \mRS}_{i^*\to kl}$ are defined only
for $z<1$. We define the full splitting functions through the relation
\begin{equation}
P_{i\to k l}^{\mRS}(z) \equiv
\frac{(1-z)}{(1-z)_+} P_{i^*\to k l}^{<\, \mRS}(z) + 
\delta_{ik}\, \gamma_\mRS(a_i)\, \delta(1-z) \, ,
\label{FullP}
\end{equation}
where we made use of the standard $+$prescription.  The factors
$\gamma_\mRS(a_i)$ and thus $P^\mRS_{i\to k l}$ are determined by
requiring that the momentum sum rules
\begin{eqnarray}
&& \int_0^1 dz\, z \left[P^{\mRS*}_{g\to g g}(z) 
   + 2 N_F\, P^{\mRS*}_{g\to q\bar{q}}(z)\right] = 0 ,
\label{sumrule_g} \\
&& \int_0^1 dz\, z \left[P^{\mRS*}_{q\to q g}(z) 
   +  P^{\mRS*}_{q\to g q}(z)\right] = 0 ,
\label{sumrule_q}
\end{eqnarray}
are satisfied in all the schemes \CDR, \HV, \FDH, i.e.\ also taking
into account terms of higher-order in $\epsilon$  if appropriate.
\Eqn{sumrule_g} determines $\gamma_{\mRS*}(g)$ and \Eqn{sumrule_q}
determines $\gamma_{\mRS*}(q)$. For \DRED, the sum rules given in
\Eqns{sumrule_g}{sumrule_q} have to be generalized in an obvious way,
since we also have to take into account $\gtilde$:
\begin{eqnarray}
&& \int_0^1 dz\, z \left[P^{\mDRED}_{\ghat\to \ghat \ghat}(z) 
   + P^{\mDRED}_{\ghat\to\gtilde \gtilde}(z)
   + 2 N_F\, P^{\mDRED}_{\ghat\to q\bar{q}}(z)\right] = 0 ,
\label{sumrule_ghat} \\
&& \int_0^1 dz\, z \left[P^{\mDRED}_{q\to q g}(z) 
   +  P^{\mDRED}_{q\to g q}(z)\right] =  
\nonumber \\
&&  \int_0^1 dz\, z \left[P^{\mDRED}_{q\to q \ghat}(z) 
   +  P^{\mDRED}_{q\to q \gtilde}(z) 
   +  P^{\mDRED}_{q\to \ghat q}(z) 
   + P^{\mDRED}_{q\to \gtilde q}(z)\right] = 0,
\label{sumrule_qdr}\\
&& \int_0^1 dz\, z \left[P^{\mDRED}_{\gtilde\to \gtilde \ghat}(z) 
   + P^{\mDRED}_{\gtilde\to\ghat \gtilde}(z)
   + 2 N_F\, P^{\mDRED}_{\gtilde\to q\bar{q}}(z)\right] = 0  .
\label{sumrule_gtilde}
\end{eqnarray}
As before, \Eqns{sumrule_ghat}{sumrule_qdr} determine
$\gamma_\mDRED(\ghat)$ and $\gamma_\mDRED(q)$ respectively, while
\Eqn{sumrule_gtilde} determines $\gamma_\mDRED(\gtilde)$. The results
are
\begin{align}
\gamma_\mCDR(g) &= \frac{11\, N_c}{6} - \frac{(3 D-8)\, T_F N_F}{3(D-2)},
&\!\! \gamma_\mCDR(q)& = \frac{(10-D)\, C_F}{4},
\label{DgammaCDR}\\
\gamma_\mFDH(g) &= \frac{(6 D-13)\, N_c}{3 (D-2)} 
  - \frac{(3 D-8)\, T_F N_F}{3(D-2)},
&\!\! \gamma_\mFDH(q) &= \frac{3 C_F}{2},
\label{DgammaFDH}\\
\gamma_\mDRED(\ghat) &= \gamma_\mFDH(g), \quad 
&\!\! \gamma_\mDRED(q)&= \gamma_\mFDH(q),
\label{DgammaDRED}
\\
\gamma_\mDRED(\gtilde) &= 2 N_c -T_F N_F.
&&
\end{align}
and $\gamma_\mHV(a_i) = \gamma_\mCDR(a_i)$.  Expanding these results
and taking into account all terms to ${\cal O}(\eps)$ we obtain the
results given in \Eqn{gammaALL}. We note that these results are also
consistent with the quark-number conservation sum rule as well as with
\Eqns{outColl}{outCollDRED}. In \FDH\ and  \DRED\ they also satisfy
the supersymmetric relation
\begin{eqnarray}
P^\mFDH_{g\to gg}+ 2 N_F\, P^\mFDH_{g\to q \bar{q}} &=&
P^\mFDH_{q\to g q} + P^\mFDH_{q\to q g}\ ,
\label{FDHsusyward} \\
P^\mDRED_{\gth\to gg}+ 2 N_F\, P^\mDRED_{\gth\to q \bar{q}} &=&
P^\mDRED_{q\to g q} + P^\mDRED_{q\to q g}\ ,
\label{DRsusyward}
\end{eqnarray}
if we set $N_c = C_F = 2\, T_F N_F$. 

Finally we mention that the \DRED\ splitting function used in
\Eqns{inCollDRED}{NonMScollDRED} are defined as
\begin{eqnarray}
P_{g\to \ghat g}^{\mDRED} &\equiv& 
  \frac{\omega_\mDRED(\ghat)}{\omega_\mDRED(g)}\,
       P_{\ghat\to \ghat \ghat}^{\mDRED} + 
  \frac{\omega_\mDRED(\gtilde)}{\omega_\mDRED(g)}\,
     P_{\gtilde\to \ghat \gtilde}^{\mDRED}\ ,
\label{PdredH}
\\
P_{g\to \gtilde g}^{\mDRED} &\equiv& 
  \frac{\omega_\mDRED(\ghat)}{\omega_\mDRED(g)}\,
     P_{\ghat\to \gtilde \gtilde}^{\mDRED} + 
  \frac{\omega_\mDRED(\gtilde)}{\omega_\mDRED(g)}\,
     P_{\gtilde\to \gtilde \ghat }^{\mDRED}\ .
\label{PdredT}
\end{eqnarray}
Using the explicit results above we find
\begin{eqnarray}
P_{g\to \ghat* g}^{<\, \mDRED} &=& 
2 N_c\, \left(\frac{D-2}{2}\frac{z}{1-z} + \frac{1-z}{z} + 
\frac{D-2}{2} z (1-z)\right)\ ,
\label{GtogGdr} \\
P_{g\to \gtilde* g}^{<\, \mDRED} &=& 
2 N_c\,\frac{4-D}{2} \left(\frac{z}{1-z} + z (1-z)\right)\ ,
\label{GtopGdr} \\
P_{q\to \ghat* q}^{<\, \mDRED} &=& 
C_F\, \left(\frac{2(1-z)}{z}+\frac{D-2}{2} z \right)\ ,
\\
P_{q\to \gtilde* q}^{<\, \mDRED} &=& 
C_F\, \frac{4-D}{2}\ z
\ ,
\end{eqnarray}
for the splitting functions used in \Eqn{InCollLimDRED}.

\section{Examples} \label{app:examples}

\subsection{$g g \to q \bar{q}$}

The process $g g \to q \bar{q}$ with massless quarks has been computed
long ago at one loop~\cite{Ellis:1985er} in \CDR\ and was one of the
processes used to determine the relations between the \HV\ and \FDH\
scheme~\cite{Kunszt:1993sd}. The one-loop matrix elements ${\cal
M}^{(1)}_{\mRS *}(g,g; q,\bar{q})$ were found to be related as given
in \Eqn{Mvirtual*}. This and the related process with massive quarks
was also at the centre of claims regarding problems with factorization
in \DRED~\cite{Beenakker:1988bq, Beenakker:1996dw, Smith:2004ck}. The
factorization issue related to the real corrections for these
processes has been solved in Ref.~\cite{Signer:2005iu}. Here we focus
on some aspects related to issues with \DRED, starting with the
virtual corrections.

\begin{figure}[t]
\begin{center}
\begin{picture}(340,120)(-30,0)
\SetOffset(0,60)
\Text(0,30)[]{${\cal M}^{(1,\rm ct)}_{\mDRED}(\ghat,\gtilde;q,\bar{q})=$ }
\SetOffset(0,0)
\Text(0,30)[]{${\cal M}^{(1,\rm ct)}_{\mDRED}(\gtilde,\gtilde;q,\bar{q})=$ }
\SetOffset(40,60)
\ZigZag(10,50)(50,50){2}{6}
\DashLine(10,10)(50,10){2.}
\ArrowLine(90,10)(50,10)
\ArrowLine(50,10)(50,50)
\ArrowLine(50,50)(90,50)
\GCirc(50,10){3}{0.9}
\GCirc(50,50){3}{0.5}
\SetOffset(140,60)
\ZigZag(10,50)(50,10){2}{10}
\DashLine(10,10)(50,50){2.}
\ArrowLine(90,10)(50,10)
\ArrowLine(50,10)(50,50)
\ArrowLine(50,50)(90,50)
\GCirc(50,10){3}{0.5}
\GCirc(50,50){3}{0.9}
\SetOffset(240,60)
\ZigZag(10,50)(30,30){2}{5}
\DashLine(10,10)(30,30){2.}
\DashLine(30,30)(70,30){2.}
\ArrowLine(90,10)(70,30)
\ArrowLine(70,30)(90,50)
\GCirc(30,30){3}{0.5}
\GCirc(70,30){3}{0.9}
\SetOffset(40,00)
\DashLine(10,50)(50,50){2.}
\DashLine(10,10)(50,10){2.}
\ArrowLine(90,10)(50,10)
\ArrowLine(50,10)(50,50)
\ArrowLine(50,50)(90,50)
\GCirc(50,10){3}{0.9}
\GCirc(50,50){3}{0.9}
\SetOffset(140,00)
\DashLine(10,50)(50,10){2.}
\DashLine(10,10)(50,50){2.}
\ArrowLine(90,10)(50,10)
\ArrowLine(50,10)(50,50)
\ArrowLine(50,50)(90,50)
\GCirc(50,10){3}{0.9}
\GCirc(50,50){3}{0.9}
\SetOffset(240,00)
\DashLine(10,50)(30,30){2.}
\DashLine(10,10)(30,30){2.}
\ZigZag(30,30)(70,30){2}{6}
\ArrowLine(90,10)(70,30)
\ArrowLine(70,30)(90,50)
\GCirc(30,30){3}{0.5}
\GCirc(70,30){3}{0.5}
\end{picture}
\end{center}
\ccaption{}{Ultraviolet counterterm diagrams due to coupling
  renormalization 
${\cal M}^{(1,\rm ct)}_{\mDRED}(\ghat,\gtilde;q,\bar{q})$ (upper line)
and ${\cal M}^{(1,\rm ct)}_{\mDRED}(\gtilde,\gtilde;q,\bar{q})$ (lower
line). Dark vertices represent counterterms $\delta Z_g^\mDRED$ and
bright vertices stand for counterterms $\delta \widetilde Z_g^\mDRED$.
\label{fig:ggqq}}
\end{figure}

The calculation of the one-loop diagrams is straightforward and we
stress once more that there is no need to disentangle $g$ into $\ghat
+\gtilde$ in the explicit calculation of the one-loop diagrams. The
only issue in the computation of ${\cal
M}^{(1)}_\mDRED(g,g,q,\bar{q})$ is renormalization. For massless
quarks we only have to consider coupling renormalization. In \CDR,
\HV\ and \FDH\ this simply amounts to adding the counterterm
\begin{equation}
\label{qq:Uvnormal}
{\cal M}^{(1)}_{\mRS *}(g,g;q,\bar{q}) = 
\bar{\cal M}^{(1)}_{\mRS *}(g,g;q,\bar{q}) +
2\, \delta  Z_g^{\mRS *}\, {\cal M}^{(0)}_{\mRS *}(g,g;q,\bar{q})\, ,
\end{equation}
where $Z_g^{\mRS *}$ is the \RS\ dependent coupling renormalization
factor (in the \MSbar\ scheme). In \DRED\ we have to split the
counterterm contributions as
\begin{eqnarray}
{\cal M}^{(1,\rm ct)}_{\mDRED}(g,g;q,\bar{q}) &=&
{\cal M}^{(1,\rm ct)}_{\mDRED}(\ghat,\ghat;q,\bar{q}) +
{\cal M}^{(1,\rm ct)}_{\mDRED}(\ghat,\gtilde;q,\bar{q}) 
\\ \nonumber &+&
{\cal M}^{(1,\rm ct)}_{\mDRED}(\gtilde,\ghat;q,\bar{q}) +
{\cal M}^{(1,\rm ct)}_{\mDRED}(\gtilde,\gtilde;q,\bar{q}) 
\end{eqnarray}
and renormalize all four parts on the r.h.s. separately. For this we
need the coupling renormalization factors $Z_g^\mDRED$ for the $\ghat
q\bar{q}$ coupling and $\widetilde Z_g^\mDRED$ for the $\gtilde
q\bar{q}$ coupling. They are well known~\cite{Jack:1993ws} and
read\footnote{%
We have set  $\alpha_s=\alpha_e$ in these results. This is allowed
since we are   working at one loop.}
\begin{eqnarray}
\delta Z_{g}^\mDRED &=& \frac{\alpha_s}{4\pi} \frac{c_\Gamma}{\eps}\, 
  \frac{(-11+\eps) N_c + 4\, T_F N_F}{6}\ ,
\label{qq:Zg4}
\\
\delta\widetilde{Z}_{g}^\mDRED &=& 
  \frac{\alpha_s}{4\pi} \frac{c_\Gamma}{\eps}\, 
  \left( \frac{1}{2\, N_c} - \frac{3\, N_c}{2} + 
   T_F N_F + \eps\, {\rm Finite} \right)\ .
\label{qq:Zgtil}
\end{eqnarray}
The finite ${\cal O}(\alpha_s)$ term in \Eqn{qq:Zg4} is required
because we use the \MSbar\ and not the \DRbar\ scheme. The divergent
part of \Eqn{qq:Zgtil} is determined by requiring the cancellation of
UV singularities in the off-shell $\gtilde\gtilde$ Green function. The
finite ${\cal O}(\alpha_s)$ terms in \Eqn{qq:Zgtil} would have to be
determined by a renormalization scheme. However, they will not affect
the final result. As is to be expected, this allows us to perform the
calculation without specifying a renormalization scheme for the
unphysical gluons.   The counterterm ${\cal M}^{(1,\rm
  ct)}_{\mDRED}(\ghat,\ghat;q,\bar{q})$ is simply given by $2 \delta
Z_{g}^\mDRED\, {\cal M}^{(0)}_{\mDRED}(\ghat,\ghat;q,\bar{q})$. As
illustrated in Figure~\ref{fig:ggqq}, the counterterm ${\cal
  M}^{(1,\rm ct)}_{\mDRED}(\ghat,\gtilde;q,\bar{q})$ is given by
$(\delta Z_{g}^\mDRED + \delta\widetilde{Z}_{g}^\mDRED) {\cal
  M}^{(0)}_{\mDRED}(\ghat,\gtilde;q,\bar{q})$, while ${\cal M}^{(1,\rm
  ct)}_{\mDRED}(\gtilde,\gtilde;q,\bar{q})$ is not proportional to the
corresponding tree-level amplitude. We explicitly verified that after
renormalization the one-loop matrix element
\begin{equation}
{\cal
M}^{(1)}_{\mDRED}(g,g;q,\bar{q})=\bar{\cal
M}^{(1)}_{\mDRED}(g,g;q,\bar{q})+{\cal
M}^{(1,\rm ct)}_{\mDRED}(g,g;q,\bar{q})
\end{equation}
in \DRED\ is
related to the other schemes as given in  \Eqn{MVfhd-dred}.

The calculation of the real matrix elements is trivial. They are
simply the four-dimensional results, \Eqn{Mtree}, and the
corresponding real cross section can be obtained in \DRED\  by
integrating these matrix elements over the phase space. The only
remaining and main issue is the factorization of the initial state
collinear singularities. According to our discussion, \Eqn{collctX},
it is clear that we will have to add
\begin{eqnarray}
\label{ggqq:cctG}
\lefteqn{d\sigma^{\rm coll}_{\mDRED,\mMS}(g,g;q,\bar{q}) = } && 
\\ \nonumber &&
\frac{\alpha_s}{2\pi} \frac{c_\Gamma}{\eps}
\int dz\, \Big[
\Big( P_{g\to \ghat g}^{\mDRED}(z) 
        - P_{g\to g g}^{\mMS\, \eps}(z)\Big)
 \, d\sigma^{(0)}_\mDRED(\ghat(z\, p_1),g(p_2); q, \bar{q})
\\ \nonumber
&& \qquad \qquad +\  \Big( P_{g\to q \bar{q}}^{\mDRED}(z)  
        - P_{g\to q \bar{q}}^{\mMS\, \eps}(z)\Big)
 \, d\sigma^{(0)}_\mDRED(q(z\, p_1),g(p_2); g, q )
\\ \nonumber
&&\qquad \qquad +\  \Big( P_{g\to \bar{q} q}^{\mDRED}(z)  
        - P_{g\to \bar{q} q}^{\mMS\, \eps}(z)\Big)
 \, d\sigma^{(0)}_\mDRED(\bar{q}(z\, p_1),g(p_2);\bar{q} ,g ) 
\\ \nonumber
&&
\qquad \qquad +\  P_{g\to \gtilde g}^{\mDRED}(z) 
 \, d\sigma^{(0)}_\mDRED(\gtilde(z\, p_1),g(p_2); q, \bar{q} )\Big]
 + \{ 1 \leftrightarrow 2\} \ .
\end{eqnarray}
The conversion to the \MSbar\ scheme requires the terms
\begin{eqnarray}
P_{g\to g g}^{\mMS\, \eps} &\equiv& 
 P_{g\to g g}^{\mCDR}
- \left[P_{g\to g g}^{\mCDR} \right]_{D\to 4}
= \eps \, \frac{T_F\, N_F}{3} \, \delta(1-z) ,
\label{ggh:PgggConv}\\
P_{g\to q \bar{q}}^{\mMS\, \eps} &\equiv&
 P_{g\to q \bar{q}}^{\mCDR}
- \left[P_{g\to q \bar{q}}^{\mCDR} \right]_{D\to4}
= - \eps \, T_F \,2\, z (1-z) .
\label{ggh:PgqqConv}
\end{eqnarray}
It is the last term on the r.h.s. of \Eqn{ggqq:cctG} that is
non-standard and deserves special mention since it resolves the issue
regarding the seemingly non-factorizing corrections in \DRED.

Let us add a comment on why the factorization problem of Refs.\
\cite{Beenakker:1988bq, Beenakker:1996dw, Smith:2004ck} was found in
the context of the process with massive quarks, discussed below,
rather than the one with massless quarks. The reason is the fact that
in the present, massless case, the \DRED\ cross sections for the
$\gtilde$ and $\ghat$ initial states happen to be equal,
\begin{equation}
d\sigma^{(0)}_\mDRED(\ghat( p_1),g(p_2); q, \bar{q})
=
d\sigma^{(0)}_\mDRED(\gtilde(p_1),g(p_2); q, \bar{q})
.
\label{DREDrelation}
\end{equation}
Hence, in \Eqn{ggqq:cctG} one can combine the terms in the first and
the last line to 
\begin{equation}
P_{g\to gg}^{\mDRED}(z)\,
d\sigma^{(0)}_\mDRED(g(z p_1),g(p_2); q, \bar{q}),
\end{equation}
and the process is seen to factorize even without distinguishing
between $\ghat$ and $\gtilde$.

\subsection{$g g \to Q \bar{Q}$}

Problems to reconcile factorization with \DRED\ were first mentioned
in the context of this process with massive final state quarks
\cite{Beenakker:1988bq}.  As explained in
Ref.~\cite{Signer:2005iu} and the present paper, the factorization
problem disappears if $\ghat$ and $\gtilde$ are treated as separate
partons in formulas such as \Eqns{InCollLimDRED}{inCollDRED} or in the
last line of \Eqn{ggqq:cctG}. The reason why the apparent problem has
been found only in the massive process $gg\to Q\bar{Q}$ is not related
to quark masses but to \Eqn{DREDrelation}, which happens to hold in
the massless case.

With this in mind, the massive process can be treated in the same way
as the massless one. The only additional complication in the case of
massive quarks is to consider the \RS\ dependence of the mass
renormalization $Z_m^\mRS$ and external wave-function renormalization
$Z_Q^\mRS$ for massive quark lines. The \RS\ dependence of these
renormalization factors has been considered before (see
e.g. Ref.~\cite{Harris:2002md}) and, using for example the pole scheme
to define the mass of the heavy quark $m$, can be summarized as
follows:
\begin{eqnarray}
Z_m^\mCDR &=& Z_m^\mHV = 1 + 
  \frac{\alpha_s}{4\pi} \frac{c_\Gamma}{\eps} \, C_F\, 
  \left(\frac{m^2}{\mu^2}\right)^{-\eps} 
  \left( - \frac{3}{\eps} -4\right)\ ,
\label{ggqq:ZmCDR} 
\\
Z_m^\mDRED &=& Z_m^\mFDH = 1 + 
  \frac{\alpha_s}{4\pi} \frac{c_\Gamma}{\eps} \, C_F\, 
  \left(\frac{m^2}{\mu^2}\right)^{-\eps} 
  \left( - \frac{3}{\eps} -5\right)\ ,
\label{ggqq:ZmDRED} 
\end{eqnarray}
and $Z_Q^\mRS = Z_m^\mRS$. As mentioned in the main text, $\gamma(Q)=
C_F$ is \RS\ independent~\cite{Catani:2000ef}. We have verified by
explicit calculation that using the \RS~dependent coupling
renormalization and \Eqns{ggqq:ZmCDR}{ggqq:ZmDRED} for the mass
counterterms and wave-function renormalization, the \RS\ dependence of
the virtual corrections take the form as given in \Eqn{Mvirtual}. To
use \DRED\ throughout in the calculation of this process we simply have
to use the correct \RS\ dependent collinear counterterm as given in
\Eqn{ggqq:cctG} and fold the hard partonic cross sections with
the standard PDF in the \MSbar\ factorization scheme.

\subsection{$g g \to h$}

In this example we consider the production of a Higgs $h$ through
gluon fusion in a hadronic collision. While this process is relatively
simple at one-loop it is complicated enough to illustrate all main
points discussed in the main text. The interaction of the Higgs with
gluons is given by the Lagrangian
\begin{equation}
{\cal L}_I =\frac12
 g_h\,  h\, {\rm tr}\left(F^{\mu\nu}F_{\mu\nu}\right)\, ,
\end{equation}
where the coupling $g_h$ has mass dimension $-1$. In \DRED\ we have
to distinguish between the coupling for $\ghat \ghat h$, denoted by
$g_h$ and the coupling for $\gtilde \gtilde h$, denoted by
$\gtilde_h$. At tree level the two couplings are the same, but they
differ at higher orders.

The only process that contributes at tree level is $g(p_1)\,g(p_2)\to
h$.  The matrix elements are given by
\begin{equation}
{\cal M}^{(0)}_\mRS(\gth,\gth;h) = 
g_h^2\, \omega_\mRS(\gth)\, \frac{s_{12}^2}{4} 
\end{equation}
with $\gth = g$ in \CDR,\HV\ and \FDH\ and $\gth \in
\{\ghat,\gtilde\}$ for \DRED\ and $ {\cal
  M}^{(0)}_\mDRED(\gtilde,\ghat;h) = 0$.

For the calculation of the ${\cal O}(\alpha_s)$ corrections to $g\, g
\to h$ the distinction between $g_h$ and $\gtilde_h$ at tree level
will be relevant for the renormalization.  Importantly, however, for
the non-trivial part of the explicit calculation of the virtual and
real corrections we can set $g_h =\gtilde_h$ and we do not have to
distinguish between $\ghat$ and $\gtilde$ in loop diagrams.

First we discuss the virtual corrections and how their
\RS\ dependences arise. The explicit calculation of the two
non-vanishing one-loop diagrams in Feynman gauge results in the
following unrenormalized one-loop matrix elements:
\begin{equation}
\label{ggh:M1bargg}
{\cal \bar{M}}^{(1)}_\mRS(\gth_1,\gth_2;h) = {\cal
M}^{(0)}_\mRS(\gth_1,\gth_2;h)\, \frac{\alpha_s}{2\pi}\, c_\Gamma\,
\left(- \frac{2N_c}{\eps^2} \right) 
\left|-\frac{s_{12}}{\mu^2}\right|^{-\eps}
\end{equation}
where
\begin{equation}
\label{ggh:sreal}
\left|-\frac{s_{12}}{\mu^2}\right|^{-\eps} \equiv
{\rm Re}\! \left(-\frac{s_{12}}{\mu^2}\right)^{-\eps} = 
 \left(\frac{s_{12}}{\mu^2}\right)^{-\eps} -
\frac{\eps^2\, \pi^2}{2} + {\cal O}(\eps^3).
\end{equation}
In order to obtain the counterterms we only need to perform a
renormalization transformation of the couplings, $g_h\to Z_{gh}g_h$,
$\gtilde_h\to \widetilde{Z}_{gh}\gtilde_h$. We use the \MSbar\ scheme
to define the renormalization constants in all \RS. The results
read
\begin{eqnarray}
&& Z_{gh}^\mCDR = Z_{gh}^\mHV = 
1 + \frac{\alpha_s}{4\pi} \frac{c_\Gamma}{\eps}\, 
  \frac{-11 N_c + 4\, T_F N_F}{3}\ ,
\label{ggh:Zghd}
\\
&& Z_{gh}^\mDRED = Z_{gh}^\mFDH = 
1 + \frac{\alpha_s}{4\pi} \frac{c_\Gamma}{\eps}\, 
  \frac{(-11+\eps) N_c + 4\, T_F N_F}{3}\ ,
\label{ggh:Zgh4}
\\
&& \widetilde{Z}_{gh}^\mDRED = 
1 + \frac{\alpha_s}{4\pi} \frac{c_\Gamma}{\eps}\, 
  \left(-4\, N_c + 2\, T_F N_F + \eps\, {\rm Finite} \right)\ .
\label{ggh:Zghtil}
\end{eqnarray}
As expected, $g_h$ renormalizes like the square of the strong
coupling, and the difference between $Z_{gh}^\mCDR$ and
$Z_{gh}^\mDRED$ is in agreement with the corresponding scheme
difference of $\alpha_s$~\cite{Altarelli:1980fi,Kunszt:1993sd}, see
also Ref.~\cite{Jack:1993ws}. Writing
$Z_{gh}^{\mRS} = 1 + \delta Z_{gh}^{\mRS}$ we have
\begin{equation}
{\cal M}^{(1)}_{\mRS*}(g,g;h) =
{\cal \bar{M}}^{(1)}_{\mRS*}(g,g;h) +
2\ \delta Z_{gh}^{\mRS*}\,  {\cal M}^{(0)}_{\mRS*}(g,g;h)
\label{ggh:Mren*}
\end{equation}
for \CDR, \HV\ and \FDH.
In the case of \DRED\ we obtain
\begin{eqnarray}
{\cal M}^{(1)}_{\mDRED}(g,g;h) 
\label{ggh:MrenDred}
&=&
{\cal \bar{M}}^{(1)}_{\mDRED}(\ghat,\ghat;h) +
2\ \delta Z_{gh}^{\mDRED}\,  {\cal M}^{(0)}_\mDRED(\ghat,\ghat;h)
\\ 
\nonumber &+&
{\cal \bar{M}}^{(1)}_{\mDRED}(\gtilde,\gtilde;h) +
2\ \delta \widetilde{Z}_{gh}^{\mDRED}\,  
{\cal M}^{(0)}_\mDRED(\gtilde,\gtilde;h)\ .
\end{eqnarray}
Neglecting terms of ${\cal O}(\eps)$ this entails
\begin{equation}
{\cal M}^{(1)}_{\mRS}(g,g;h) =  \frac{\alpha_s\, c_\Gamma}{2\pi}
{\cal M}^{(0)}_\mRS(g,g;h)\left(  
- \frac{2\, N_c}{\eps^2}   \left|-\frac{s_{12}}{\mu^2}\right|^{-\eps}\! 
- \frac{\beta_0}{\eps} + \Delta^{\rm virt}_\mRS \right)
\label{ggh:M1rs}
\end{equation}
with $\Delta^{\rm virt}_\mCDR=\Delta^{\rm virt}_\mHV=0$, $\Delta^{\rm
  virt}_\mFDH=N_c/3$ and $\Delta^{\rm virt}_\mDRED=2 N_F T_F/3$.
Thus, after adding the counterterms, these expressions are in
agreement with the general formula 
\Eqn{Mvirtual}, with the finite \RS\ independent part given by ${\cal
  M}^{(1)}_\mNS(g,g;h) = g_h^2\, (N_c^2-1)\, N_F \, s_{12}^2/6$.
The scheme dependences also exemplify the formulas discussed in
Section~\ref{sec:SchemeRelations}.

Turning to the calculation of the real corrections, we also have to
take into account the processes with (anti)quarks in the initial
state. The corresponding matrix elements in \CDR\ are given by
\begin{eqnarray}
\label{ggh:mgghg}
\lefteqn{{\cal M}^{(0)}_\mCDR(g,g;h,g) = } &&
\\  &&
g_h^2\, 4\pi\alpha_s\, N_c (N_c^2-1)
\sum_{\rm cycl} \left( \frac{
(D-2) (s_{12}-s_{14})^2(s_{12}-s_{24})^2}{s_{12} s_{24} s_{14}} 
- 4 s_{12} \right)\, ,
\nonumber
\\
\label{ggh:mgqhq}
\lefteqn{ {\cal M}^{(0)}_\mCDR(g,q;h,q)=} && 
\\&&
g_h^2\, 4\pi\alpha_s\, T_F (N_c^2-1)
\frac{(D-2)(s_{12}^2+s_{14}^2) -2(D-4) s_{12} s_{14}}{2\, s_{24}} ,
\nonumber
\\
\label{ggh:mqqhg}
\lefteqn{ {\cal M}^{(0)}_\mCDR(\bar{q},q;h,g)=} && 
\\&&
g_h^2\, 4\pi\alpha_s\, T_F (N_c^2-1)
\frac{(D-2)(s_{24}^2+s_{14}^2) +2(D-4) s_{24} s_{14}}{2\, s_{12}} ,
\nonumber
\end{eqnarray}
where the sum in \Eqn{ggh:mgghg} is over all cyclic permutations
$\{p_1 \to p_2 \to -p_4\}$ and the corresponding matrix elements in
\HV, \FDH\ and \DRED\ can be obtained by setting $D\to
4$. \Eqns{ggh:mgqhq}{ggh:mqqhg} are related by crossing.

We have now all the necessary matrix elements at hand to compute cross
sections. In this example we restrict ourselves to the total cross
section $\sigma$. The cancellation of singularities and the \RS\
dependence is the same for any infrared-finite observable.

Defining $x\equiv M_h^2/s_{12} = p_3^2/s_{12}$ we write the
leading-order cross section as
\begin{equation}
\sigma^{(0)}_\mRS(gg) \equiv
\sigma^{(0)}_\mRS(g,g;h) 
= \frac{2\pi}{s_{12}} \delta(1-x) 
\langle {\cal M}^{(0)}_\mRS(g,g;h) \rangle\ .
\label{ggh:sig0}
\end{equation}
Note that in \DRED\ we have
$\sigma^{(0)}_\mDRED(gg)=\sigma^{(0)}_\mDRED(\ghat
g)=\sigma^{(0)}_\mDRED(\gtilde g)$, analogously to
\Eqn{DREDrelation}. This is another example of the special case
discussed at the end of Ref.\ \cite{Signer:2005iu}.
Obviously, the virtual corrections can be obtained by the same formula
if we replace ${\cal M}^{(0)}_\mRS$ by ${\cal M}^{(1)}_\mRS$. Defining
the function
\begin{equation}
S^{\rm virt} \equiv \delta(1-x)\
\left(\frac{e^{\gamma_E}}{4\pi}\right)^{-\eps}\,
\left(
- \frac{2\, N_c}{\eps^2}\left(\frac{s_{12}}{\mu^2}\right)^{-\eps}\,
- \frac{\beta_0}{\eps}
+ \frac{7\, N_c}{6}\, \pi^2 \right)\ ,
\end{equation}
the explicit results read
\begin{equation}
\sigma^{\rm virt}_\mRS(gg) = \frac{\alpha_s}{s_{12}}
\langle {\cal M}^{(0)}_\mRS(g,g;h) \rangle\,
\left(S^{\rm virt} + \delta(1-x) \Delta^{\rm virt}_\mRS \right) .
\label{ggh:sigv_rs}
\end{equation}
To obtain the real corrections we parametrize the phase space with the
help of the variable $y$ defined such that
\begin{equation}
s_{14} = \frac{s_{12}}{2} (1-x) (1-y); \quad
s_{24} = \frac{s_{12}}{2} (1-x) (1+y) .
\end{equation}
The real corrections can then be written as
\begin{equation}
\label{ggh:realGG}
\sigma^{\rm real}_\mRS(a_1 a_2) =
\frac{(16 \pi)^{\eps-1}}{\Gamma(1-\eps)}\, s_{12}^{-\eps}\,
(1-x)^{1-2\eps} \int_{-1}^1dy\, (1-y^2)^{-\eps} 
\langle M^{(0)}_\mRS(a_1, a_2;h, a_3) \rangle
\end{equation}
and evaluated using  distribution identities like
\begin{eqnarray}
\frac{1}{(1-x)^{1+2\eps}} &=& - \frac{1}{2\eps}\, \delta(1-x) + 
\frac{1}{(1-x)_+} - 2\eps \left(\frac{\ln(1-x)}{1-x}\right)_+
+{\cal O}(\eps^2)
\nonumber
\\ \label{ggh:dist} 
 &\equiv& - \frac{1}{2\eps}\, \delta(1-x) + 
 I_+(x) - 2\eps\, L_+(x) +{\cal O}(\eps^2) .
\end{eqnarray}
In \CDR\ and \DRED\ the integration is straightforward, while in
\HV\ and \FDH\ the distinction between internal and external gluons as
defined in Section~\ref{sec:DR} leads to a complication. As discussed
there, in \HV\ and \FDH\ soft or collinear gluons have to be treated
not as $\gbar$ but as $\ghat$ and $g$, respectively. In order to
ensure this correct treatment, one can subtract the collinear limit of
the integrand in \Eqn{ggh:realGG} and integrate it separately,
either explicitly or using e.g.\ the dipole formalism as in
Ref.~\cite{Catani:1996pk}. Alternatively, one can split the integrand
using distribution identities for $(1+y)^{-1-\epsilon}$ similar to
\Eqn{ggh:dist} and treat each term as appropriate.

Ultimately, defining the functions
\begin{eqnarray}
\nonumber
S^{\rm real}(gg) &=& N_c\, 
\left(\frac{e^{\gamma_E}}{4\pi}\, \frac{s_{12}}{\mu^2}\right)^{-\eps} 
\bigg( \frac{2}{\eps^2} \, \delta(1-x) - 
\frac{4}{\eps} \left(I_+(x) - x (2-x+x^2) \right)
\phantom{x}
\\  &&
+\  \frac{11}{3}\, (x-1)^3 
  + 8\, L_+(x)\, \left(x^2 - x +1\right)^2
  - \frac{\pi^2}{2}\, \delta(1-x)  \bigg)\ ,
\\
\nonumber
S^{\rm real}(gq) &=&  C_F\, 
\left(\frac{e^{\gamma_E}}{4\pi}\, \frac{s_{12}}{\mu^2}\right)^{-\eps} 
\bigg( \frac{2x-2-x^2}{\eps} + \frac{6x-x^2-3}{2}
\\  && 
  -\ 2 L_+(x)\, (1-x)\left(2x-2-x^2 \right) \bigg)\ ,
\end{eqnarray}
the real cross section can be written as
\begin{eqnarray}
\sigma^{\rm real}_\mRS(gg) &=&  \frac{\alpha_s}{s_{12}}
\langle {\cal M}^{(0)}_\mRS(g,g;h) \rangle\, 
\left( S^{\rm real}(gg) + \Delta^{gg}_\mRS \right)\ ,
\label{ggh:sigGGr_rs} \\[5pt]
\sigma^{\rm real}_\mRS(gq) &=& \frac{\alpha_s}{s_{12}}
\langle {\cal M}^{(0)}_\mRS(g,g;h) \rangle
\left(S^{\rm real}(gq) + \Delta^{gq}_\mRS \right)\ ,
\label{ggh:sigGQr_rs} 
\end{eqnarray}
with 
$\Delta^{gg}_\mCDR=\Delta^{gg}_\mHV=\Delta^{gg}_\mDRED=0$,
$\Delta^{gg}_\mFDH=-4 N_c\, x^2(1-x)$,
$\Delta^{gq}_\mCDR=\Delta^{gq}_\mHV=0$ and
$\Delta^{gq}_\mFDH=\Delta^{gq}_\mDRED=-C_F\, x^2$.
The matrix element ${\cal M}^{(0)}_\mRS(\bar{q},q;h,g)$,
\Eqn{ggh:mqqhg}, will not produce any singularities upon integration
over phase space. Thus $\sigma^{\rm real}_\mRS(\bar{q}q)$ is finite.

The remaining ingredients needed for the hard partonic cross section
are the collinear counterterms suitable for the \MSbar\ factorization
schemes. They are constructed according to
\Eqn{collctX} and read
\begin{eqnarray}
\label{ggh:cctG}
\lefteqn{d\sigma^{\rm coll}_{\mMS,\mRS }(gg) = } && 
\\ \nonumber &&
\frac{\alpha_s}{2\pi} \frac{c_\Gamma}{\eps}
\int dz\, \Big[
\Big( P_{g \to\ghat g}^{\mRS}(z) 
        - P_{g\to g g}^{\mMS\, \eps}(z)\Big)
 \, d\sigma^{(0)}_\mRS(\ghat(z\, p_1),g(p_2);h)
\\ \nonumber
&& \qquad \qquad +\  P_{g\to \gtilde g}^{\mRS}(z) 
 \, d\sigma^{(0)}_\mRS(\gtilde(z\, p_1),g(p_2);h) \Big]
 + \{ 1 \leftrightarrow 2\} \ ,
\\
\label{ggh:cctQ}
\lefteqn{d\sigma^{\rm coll}_{\mMS,\mRS }(gq) = } && 
\\ \nonumber &&
\frac{\alpha_s}{2\pi} \frac{c_\Gamma}{\eps}
\int dz\, \Big[
\Big( P_{q\to \ghat q}^{\mRS}(z) 
        - P_{q\to g q}^{\mMS\, \eps}(z)\Big)
 \, d\sigma^{(0)}_\mRS(g(p_1),\ghat(z\, p_2);h)
\\ \nonumber
&& \qquad \qquad +\ P_{q\to \gtilde q }^{\mRS}(z) 
 \, d\sigma^{(0)}_\mRS(g(p_1),\gtilde(z\, p_2);h)   \Big]\ ,
\end{eqnarray}
where
\begin{align}
P_{q\to g q}^{\mMS\, \eps} &\equiv 
 P_{q\to g  q}^{\mCDR}
- \left[P_{q\to g q}^{\mCDR} \right]_{D\to4}
= -\eps\, C_F\, z
\ ,
\label{ggh:PqgqConv}
\end{align}
in addition to \Eqn{ggh:PgggConv}
is taking into account the conversion to the \MSbar\ factorization
scheme. In order to present the explicit results for the collinear
counterterms we introduce the functions
\begin{eqnarray}
S^{\rm coll}(gg) &=& \frac{c_\Gamma}{\eps}\, 
4\, N_c \left( x^2 I_+(x) + (1-x)(1+x^2)\right) 
+ \frac{c_\Gamma}{\eps}\, \beta_0\, \delta(1-x) ,
\\
S^{\rm coll}(gq) &=&  
\frac{c_\Gamma}{\eps}\, C_F\left(2 - 2 x + x^2 \right)  .
\end{eqnarray}
The collinear counterterms can then be written as
\begin{eqnarray}
\sigma^{\rm coll}_\mRS(gg) &=& \frac{\alpha_s}{s_{12}}
\langle {\cal M}^{(0)}_\mRS(g,g;h) \rangle
\left(S^{\rm coll}(gg) -\Delta^{gg}_\mRS - 
\Delta^{\rm virt}_\mRS \, \delta(1-x) \right),
\label{ggh:sigGGc_rs} \\
\sigma^{\rm coll}_\mRS(gq) &=& \frac{\alpha_s}{s_{12}}
\langle {\cal M}^{(0)}_\mRS(g,g;h) \rangle
\left(S^{\rm coll}(gq) - \Delta^{gq}_\mRS \right).
\label{ggh:sigGQc_rs} 
\end{eqnarray}
It is now easy to see that the subtracted, partonic cross sections
\begin{eqnarray}
\hat\sigma(gg) &\equiv&  \left[\sigma_\mRS^{\rm Born}(gg) + 
  \sigma_\mRS^{\rm virt}(gg) +  \sigma_\mRS^{\rm real}(gg) + 
  \sigma_\mRS^{\rm coll}(gg)  \right]_{D\to 4}\ ,
\\
\hat\sigma(gq) &\equiv& \left[ \sigma_\mRS^{\rm real}(gq) + 
  \sigma_\mRS^{\rm coll}(gq) \right]_{D\to 4}\ ,
\\
\hat\sigma(\bar{q}q) &\equiv& \left[ \sigma_\mRS^{\rm real}(\bar{q}q)
   \right]_{D\to 4}\ ,
\end{eqnarray}
are finite and \RS\ independent. 
In the sum all the \RS\ dependent $\Delta_{\mRS}$ terms drop out, and
in all \RS\ we obtain the well-known result that can be found e.g.\ in
Ref.~\cite{Dawson:1990zj}.

To summarize the calculation in \DRED: Both the one-loop diagrams and
the real corrections and phase space integrals can be computed in a
straightforward way, using only full gluons $g$. The split
$g=\ghat+\gtilde$ has to be used in the evaluation of the collinear
counterterms $d\sigma^{\rm coll}$ and in the computation of the UV
counterterms.  In the end, the cross sections
$\hat\sigma(gg),\hat\sigma(gq),\hat\sigma(q\bar{q})$, as well as
$\hat\sigma(qg),\hat\sigma(g\bar{q}),\hat\sigma(\bar{q}g),
\hat\sigma(q\bar{q})$ are \RS\ independent. The hadronic cross section
is obtained by convoluting them with the standard parton distribution
functions obtained in the \MSbar\ factorization scheme.



\end{document}